\documentclass[sigconf, nonacm]{acmart}

\newcommand\vldbdoi{XX.XX/XXX.XX}

\newcommand\vldbavailabilityurl{URL_TO_YOUR_ARTIFACTS}

\usepackage{balance}
\usepackage{subcaption}
\usepackage[linesnumbered,ruled,vlined]{algorithm2e} 

\begin{document}

\title{Structured Data Search: How far are we from ChatGPT?}
\title{Tursio Database Search: How far are we from ChatGPT?}

\author{Sulbha Jain}
\authornote{Work done while consulting for Tursio.}
\affiliation{%
  \institution{Independent Consultant}
  \city{Bellevue}
  \country{USA}
}


\author{Shivani Tripathi, Shi Qiao, Alekh Jindal}
\affiliation{%
  \institution{Tursio}
  \city{Bellevue}
  \country{USA}\vspace{0.5cm}
}









\renewcommand{\shortauthors}{Jain et al.}

\vspace{0.3cm}

\begin{abstract}
Business users need to search enterprise databases using natural language, just as they now search the web using ChatGPT or Perplexity. However, existing benchmarks---designed for open-domain QA or text-to-SQL---do not evaluate the end-to-end quality of such a search experience. We present an evaluation framework for structured database search that generates realistic banking queries across varying difficulty levels and assesses answer quality using relevance, safety, and conversational metrics via an LLM-as-judge approach. We apply this framework to compare Tursio, a database search platform, against ChatGPT and Perplexity on a credit union banking schema. Our results show that Tursio achieves answer relevancy statistically comparable to both baselines (97.8\% vs.\ 98.1\% on simple, 90.0\% vs.\ 100.0\% on medium, 89.5\% vs.\ 100.0\% on hard questions), even though Tursio answers from a structured database while the baselines generate responses from the open web. We analyze the failure modes, identify database completeness as the primary bottleneck, and outline directions for improving both the evaluation methodology and the systems under evaluation.
\end{abstract}

\maketitle



\section{Introduction}

Databases are the primary repositories for operational data in enterprises, and better access to them can unlock significant value, from improved processes to faster decision-making and enhanced customer experiences. Unfortunately, operational data is raw and uncurated: column names are often cryptic, relationships between tables are implicit, and domain-specific conventions are undocumented~\cite{ProfilingRelationalData15}. This makes it difficult for business users, who lack the technical expertise to write SQL or use text-to-SQL interfaces~\cite{bird_bench, beaver_benchmark}, to search their data effectively. Ideally, these users should be able to query databases using natural language, just as they search the web using Google or ChatGPT. While recent efforts from Databricks~\cite{databricks-genie-knowledge-store}, Snowflake~\cite{snowflake-views-semantic-overview}, and Google~\cite{google-gemini-database-understanding} have made progress in this direction, the uncurated nature of operational data continues to challenge large language models (LLMs) in understanding the database, leading to inaccurate or even hallucinated answers~\cite{sun-times-ai-misinformation, bbc-ai-mistakes}.

\begin{figure*}[t]
  \centering
  \begin{subfigure}[b]{0.24\textwidth}
    \centering
    \includegraphics[width=\textwidth]{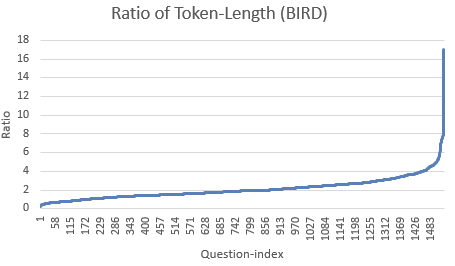}
    \caption{BIRD}
    \label{fig:ratio-bird}
  \end{subfigure}
  \hfill
  \begin{subfigure}[b]{0.24\textwidth}
    \centering
    \includegraphics[width=\textwidth]{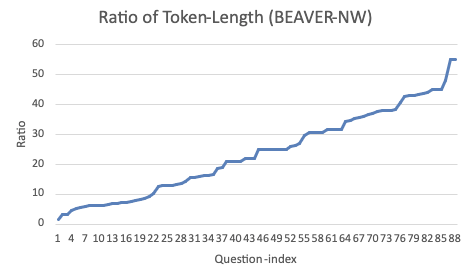}
    \caption{BEAVER-NW}
    \label{fig:ratio-beaver-nw}
  \end{subfigure}
  \hfill
  \begin{subfigure}[b]{0.24\textwidth}
    \centering
    \includegraphics[width=\textwidth]{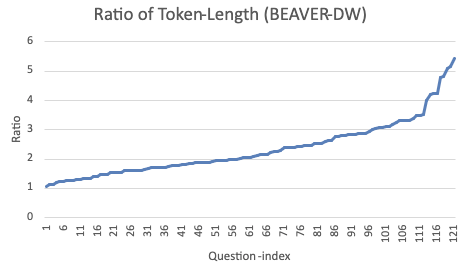}
    \caption{BEAVER-DW}
    \label{fig:ratio-beaver-dw}
  \end{subfigure}
  \hfill
  \begin{subfigure}[b]{0.24\textwidth}
    \centering
    \includegraphics[width=\textwidth]{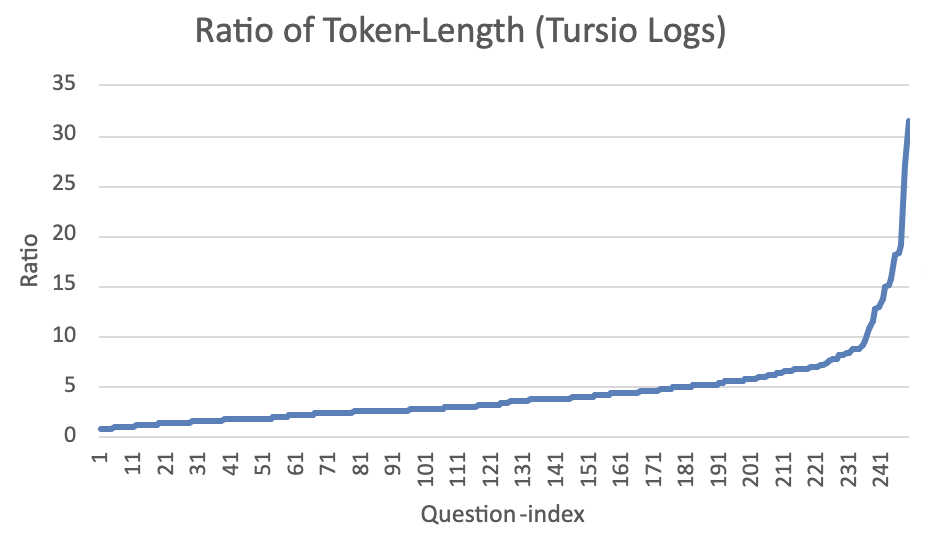}
    \caption{Tursio (production)}
    \label{fig:ratio-tursio}
  \end{subfigure}
  \vspace{-0.2cm}
  \caption{SQL-to-question token-length ratio, sorted by ratio, across benchmarks and production logs. BIRD questions are mostly literal (ratio $\approx$ 1). Enterprise workloads (BEAVER, Tursio) show significantly higher ratios, confirming that business questions require far more SQL than their surface form suggests.}
  \vspace{-0.2cm}
  \label{fig:token-ratio}
\end{figure*}

A common assumption is that natural language questions can be directly translated to SQL, as captured by the ``NL2SQL'' paradigm. In practice, however, business users ask high-level, abstract questions that omit explicit references to tables, join paths, filter conditions, or metric definitions~\cite{nl2sql-blog}. To quantify this gap, we measure the ratio of SQL token-length to question token-length across benchmarks and production logs (Figure~\ref{fig:token-ratio}). BIRD~\cite{bird_bench}, the most widely used NL2SQL benchmark, shows a largely flat curve with ratios near 1, reflecting its literal, schema-aware questions. BEAVER~\cite{beaver_benchmark}, designed to be more representative of enterprise workloads, tells a different story: its NW dataset exhibits ratios up to 55$\times$, while even its DW dataset reaches 5$\times$. Production Tursio logs confirm this pattern, with a long tail reaching 30$\times$. This demonstrates that real business questions are fundamentally different from academic NL2SQL benchmarks and require systems that can infer the implicit structure behind high-level intent.

\begin{figure}[t]
  \centering
  \includegraphics[width=\columnwidth]{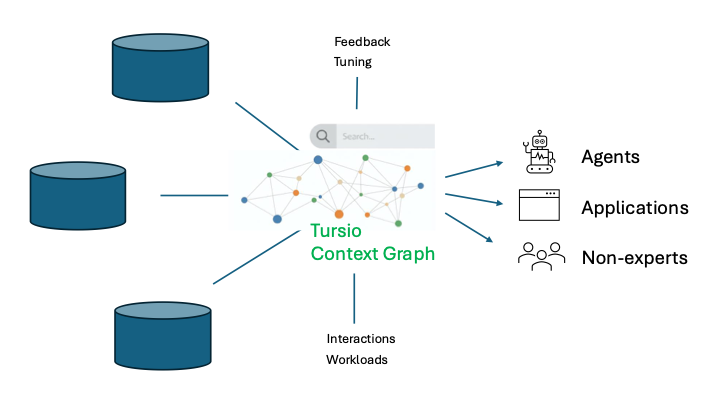}
  \vspace{-0.3cm}
  \caption{Tursio search platform connects databases through a context graph, enabling natural language search for agents, applications, and non-expert users.}
  \Description{Tursio architecture showing databases connected to a context graph with a search interface, serving agents, applications, and non-experts.}
  \label{fig:tursio-overview}
  \vspace{-0.3cm}
\end{figure}

The Tursio search platform addresses this challenge by automatically inferring a semantic knowledge graph over the database and using it as context for processing natural language queries~\cite{DatabasesSearchableDeepContext15}. As shown in Figure~\ref{fig:tursio-overview}, Tursio builds a context graph from one or more databases, incorporating feedback and workload signals, and exposes a natural language search interface to agents, applications, and non-expert users. Users can ask high-level questions---not necessarily ones directly translatable to SQL---and Tursio resolves which parts of the graph are relevant and what relationships to traverse for generating a response. We refer the readers to~\cite{DatabasesSearchableDeepContext15} for more details. The natural question, then, is whether Tursio succeeds in providing a ChatGPT-like search experience for business users.

In this paper, we evaluate Tursio's search quality compared to ChatGPT and Perplexity, the gold standards for high quality chatting and search. Specifically, we consider a realistic search scenario where credit union users ask banking-related questions over a Symitar~\cite{symitar} core banking schema. We simulate the search experience by generating queries with varying complexity, and then asking each system to return relevant results. For ChatGPT and Perplexity, we rephrase the question to be answerable from the web. The goal is to see whether Tursio provides a comparable search experience.
Our results show that Tursio achieves search relevance comparable to ChatGPT and Perplexity, matching their quality on structured data. Specifically, we make the following contributions:

\begin{itemize}
  \item We design an evaluation framework for structured database search that measures end-to-end answer quality---not just SQL correctness---including a parameterized question generation pipeline, an open-domain mapping step for fair benchmarking, and an LLM-as-judge evaluation approach with relevance, safety, and conversational metrics (Section~\ref{sec:evaluation}).
  \item We conduct the first head-to-head comparison of a structured database search system (Tursio) against leading AI web search engines (ChatGPT, Perplexity). Our results show statistically comparable answer relevancy across all difficulty levels (Section~\ref{sec:results}).
  \item We provide a detailed analysis of failure modes, identifying database completeness---and not model comprehension---as the primary bottleneck, and characterize three distinct performance profiles shaped by each system's data access model (Section~\ref{sec:analysis}).
  \item We outline concrete directions for improving both the evaluation methodology and the systems under evaluation, including multi-turn assessment, data coverage enrichment, and pipeline automation (Section~\ref{sec:future}).
\end{itemize}

\section{Evaluating Search Quality}
\label{sec:evaluation}


Evaluating search quality over structured databases presents unique challenges compared to web search evaluation. Existing benchmarks are designed for open-domain retrieval and do not capture the complexity of real enterprise queries involving joins, aggregations, and multi-table reasoning. In this section, we describe our evaluation methodology. We first discuss the limitations of current benchmarks and motivate the need for a structured-data-specific approach (\S2.1). We then detail our dataset generation pipeline, which produces realistic banking queries at varying difficulty levels (\S2.2). Next, we describe our answer evaluation framework, built on DeepEval~\cite{deepeval}, that measures relevance, bias, and conversational quality across Tursio, ChatGPT, and Perplexity (\S2.3). Finally, we present the experimental settings, including the schema, personas, and KPIs used in our evaluation (\S2.4).

\subsection{Background}

Most current benchmarks for evaluating search accuracy fall into two broad categories: open-domain factual QA and text-to-SQL evaluation. Neither is well suited for assessing search quality over structured enterprise databases.

\smallskip\textbf{Open-domain QA benchmarks.} SimpleQA~\cite{simpleqa}, introduced by OpenAI in 2024, measures short-form factual accuracy of LLMs. It contains 4,326 fact-seeking questions, each with a single, unambiguous, time-invariant answer, spanning topics such as science, politics, and art. Responses are graded on a three-tier scale---correct, incorrect, or not attempted---using an LLM-based classifier. While SimpleQA is effective for evaluating parametric knowledge recall, it tests only open-domain factual retrieval: its questions are short-form, single-answer lookups with no notion of schema, joins, aggregations, or multi-table reasoning. Search platforms such as exa.ai~\cite{exa_ai} adopt SimpleQA-style benchmarks for evaluation, but these do not capture the complexity of enterprise database search.

\smallskip\textbf{Text-to-SQL benchmarks.} BIRD~\cite{bird_bench} is the most widely used NL2SQL benchmark, containing 12,751 question-SQL pairs across 95 databases spanning 37 professional domains. It introduced execution accuracy (EX) and valid efficiency score (VES) as evaluation metrics, and its questions incorporate database values, noisy data, and external knowledge---advancing beyond earlier schema-only benchmarks like Spider. However, as we showed in Section~1 (Figure~\ref{fig:token-ratio}), BIRD's questions are largely literal translations of SQL into natural language, with SQL-to-question token-length ratios near 1. Real enterprise users ask high-level, abstract questions where this ratio can reach 30--55$\times$, as confirmed by both the BEAVER benchmark~\cite{beaver_benchmark} and production Tursio logs. Furthermore, BIRD evaluates SQL correctness rather than end-to-end search quality, i.e., whether the final answer presented to a user is relevant, complete, or useful.

\smallskip\textbf{The gap.} Neither category of benchmark addresses evaluation of search over structured databases: open-domain QA benchmarks ignore schema and data structure entirely, while text-to-SQL benchmarks assume literal, schema-aware questions and evaluate query correctness rather than answer quality. To bridge this gap, we develop an evaluation methodology that generates realistic enterprise queries at varying difficulty levels and assesses final answer quality using relevance, bias, and conversational metrics.

\subsection{Dataset Generation}

\begin{figure}[t]
  \centering
  \includegraphics[width=\columnwidth]{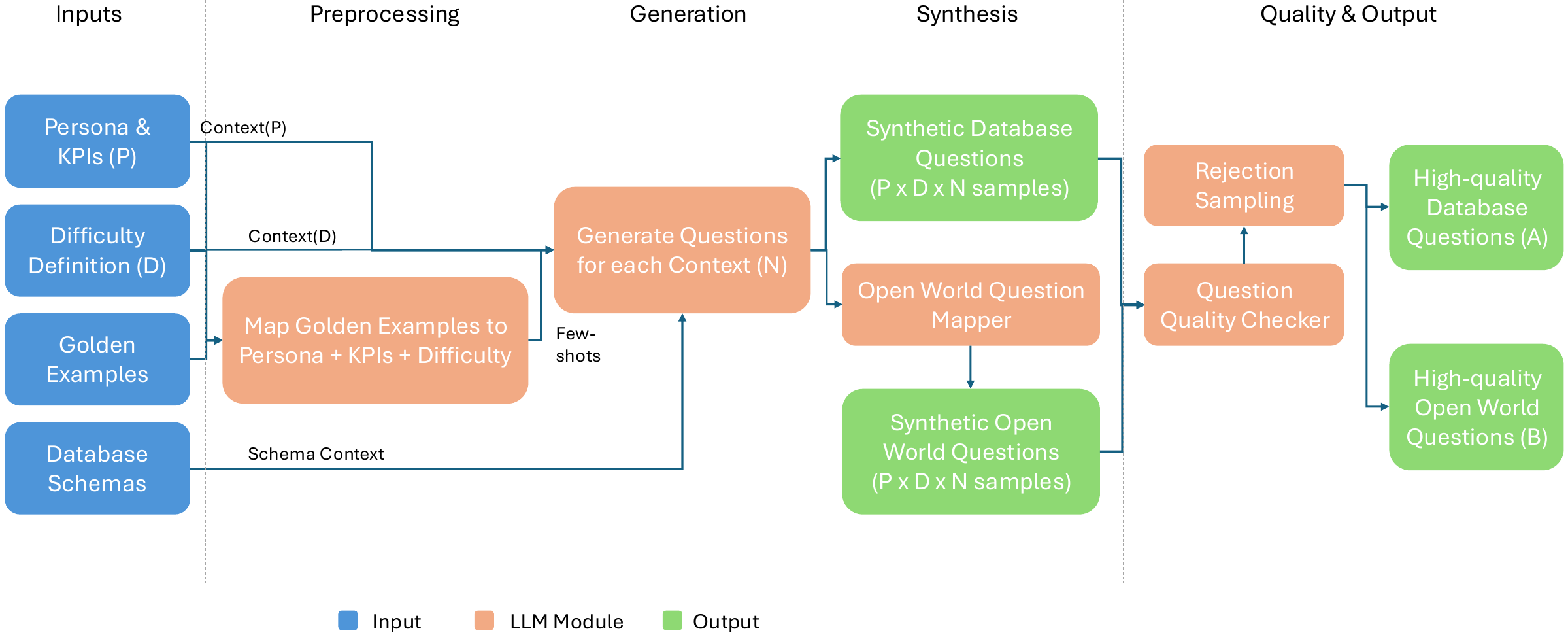}
  \caption{Question generation pipeline. Golden questions, personas, KPIs, difficulty definitions, and the database schema are combined to generate synthetic questions via an LLM. Questions are quality-checked and then mapped to real-world equivalents for benchmarking.}
  \Description{Question generation pipeline showing inputs (personas, KPIs, difficulty definitions, golden examples, DB schema) flowing through LLM-based generation, quality checking, and mapping to real-world questions.}
  \vspace{-0.2cm}
  \label{fig:question-generation}
\end{figure}

We generate a single-turn QA dataset grounded in the credit union banking domain. Starting from 15 manually curated ``golden'' questions, we systematically expand them to approximately 150 synthetic examples using the pipeline shown in Figure~\ref{fig:question-generation}. The pipeline is parameterized by three dimensions---persona ($P$), difficulty ($D$), and example questions for each context ($N$)---and produces $P \times D \times N$ candidate questions that are then quality-checked and mapped to their real-world equivalents.

\smallskip\textbf{Personas and KPIs.} We define 13 banking personas, each combining two or more roles to introduce variability in analytical focus (e.g., \emph{Risk \& Credit Analytics Manager; CRO; Compliance Officer}). Each persona is associated with relevant KPIs aligned with its typical role responsibilities. For instance, the risk-oriented persona above is mapped to KPIs such as \emph{90+ DPD Rate}, \emph{Delinquency Ratio}, and \emph{Non-Performing Loan \%}. This pairing ensures that generated questions reflect realistic analytical intent rather than simply doing generic database lookups.

\smallskip\textbf{Difficulty levels.} Questions are classified into three difficulty levels based on their analytical complexity:
\begin{itemize}
  \item \textbf{Simple:} Single-table operations involving a single metric, straightforward count, percentage, or ranking, with minimal filtering and no segmentation.
  \item \textbf{Medium:} Queries requiring filtering, grouping, or basic segmentation, possibly including a single time window or joins across a small number of tables.
  \item \textbf{Hard:} Multi-step analyses involving segmentation, time-based trends, concentration risk, or behavioral risk patterns---questions suitable for executive risk reviews or regulatory discussions.
\end{itemize}

\smallskip\textbf{Golden question segmentation.} The 15 golden questions are segmented by difficulty and mapped to relevant personas and KPIs. These annotated golden questions serve as few-shot examples for the synthetic generation step, providing the LLM with grounded context on the expected question style, complexity, and domain specificity for each persona-difficulty combination.

\smallskip\textbf{Synthetic question generation.} Using an LLM (GPT-4o-mini), we generate synthetic questions by providing the database schema, persona-KPI mappings, difficulty definitions, and sampled golden questions as few-shot examples. This produces a diverse set of questions that systematically covers the cross-product of personas and difficulty levels.

\smallskip\textbf{Mapping to real-world benchmarks.} Since the experimental schema and data are proprietary, the generated questions cannot be directly used to benchmark public systems like ChatGPT and Perplexity. To enable fair comparison, we map each synthetic question to an equivalent open-domain question using Claude Sonnet 4.5, preserving the analytical intent and difficulty while rephrasing it to be answerable using publicly available data. We used the prompt template shown in Figure~\ref{fig:question_mapper_prompt} to ensure mapping preserves the original question's intent.

\smallskip\textbf{Quality assurance.} Finally, the generated questions undergo both LLM-based and deterministic quality checks. An LLM with strong instruction-following capabilities (GPT-4) reviews questions for coherence and relevance. In addition, a suite of deterministic NLP checks is applied, including duplicate detection, semantic similarity filtering, length threshold enforcement, lexical diversity assessment, and stop-word ratio monitoring. Questions that fail any check are regenerated or discarded.

\begin{figure}[t]
\centering
\begin{minipage}{0.95\linewidth}
\scriptsize
\begin{verbatim}
You are a subject-matter expert in {DOMAIN} with deep experience
 in KPIs, personas, and data-driven analysis. 
Your responsibility is to map each original user question to the 
single best-matching question that is directly answerable using 
open web–searchable data.

Task:
For each row in the input CSV:
1. Analyze the original user question, including its analytical 
intent and scope.
2. Evaluate the candidate questions derived from web search 
results.
3. Select the ONE candidate question that most accurately 
aligns with:
   - The intent and analytical depth of the original question
   - The referenced KPI (or a clearly defined, industry-accepted 
   proxy KPI)
   - The persona’s typical decision-making needs
   - The specified difficulty level (simple / medium / hard)

Selection rules:
- Do NOT rewrite or paraphrase.
- Do NOT introduce new KPIs.
- The mapped question must be answerable using publicly available, 
open web data.
- Preserve the analytical level implied by the difficulty:
  - simple: descriptive, single-metric, snapshot questions
  - medium: comparative, segmented, or time-based questions
  - hard: multi-step, trend-based, segmented, or contextualized 
  questions
- If multiple candidates are plausible:
  - Prefer the most specific, unambiguous, and commonly used 
  industry question.
  - Prefer exact KPI alignment over loosely related metrics.
- If no candidate adequately matches all criteria, 
return an empty string ("") for "mapped_question".

Input:

Row context:
- original_question: {original_question}
- persona: {persona}
- kpi: {kpi}
- difficulty: {difficulty}

Output:
Return results STRICTLY in the following JSON format:

{{
  "original_question": "{original_question}",
  "mapped_question": "string"
}}

Output ONLY valid JSON.
Do NOT include explanations, comments, or any additional text.
\end{verbatim}
\end{minipage}
\vspace{-0.2cm}
\caption{Prompt template to map questions to real-world equivalents.}
\label{fig:question_mapper_prompt}
\vspace{-0.3cm}
\end{figure}

\subsection{Answer Evaluation}

\begin{figure}[t]
  \centering
  \includegraphics[width=\columnwidth]{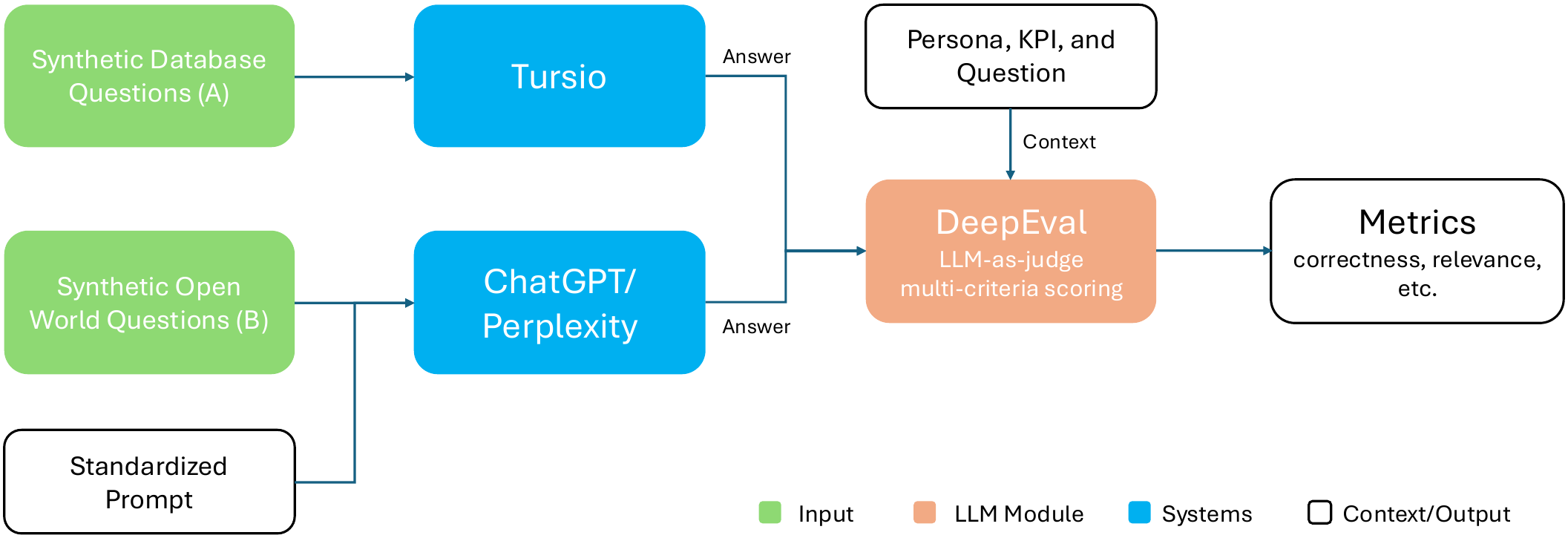}
  \vspace{-0.2cm}
  \caption{Answer evaluation pipeline. Synthetic custom questions are answered by Tursio, while their real-world equivalents are answered by ChatGPT and Perplexity. All responses are evaluated by DeepEval using the question, persona, and KPI as context.}
  \Description{Answer evaluation pipeline showing Tursio answering custom questions and ChatGPT answering real-world questions, with both sets of answers evaluated by DeepEval.}
  \vspace{-0.3cm}
  \label{fig:answer-eval}
\end{figure}

Given the generated questions, we produce answers from three systems and evaluate them using automated metrics. Figure~\ref{fig:answer-eval} illustrates the end-to-end evaluation pipeline.

\smallskip\textbf{Answer generation.} For each question, answers are generated from three platforms: (i)~Tursio, which queries the structured credit union database; (ii)~ChatGPT (as of December 2025); and (iii)~Perplexity (as of December 2025). Since ChatGPT and Perplexity do not have access to the proprietary database, they receive the real-world equivalent questions produced by the mapping step described in \S2.2, which are answerable using public data. All systems use a standardized prompt template that specifies the persona context and constrains responses to 3--5 sentences, ensuring comparable output across platforms.

\smallskip\textbf{Evaluation framework.} We evaluate answer quality using DeepEval~\cite{deepeval}, an open-source LLM evaluation framework built on the G-Eval methodology~\cite{geval}. DeepEval uses an LLM-as-judge (GPT-4.1) to score responses on a $[0, 1]$ scale across multiple dimensions and supports both built-in and custom metrics. We use the following metrics in our evaluation:

\begin{itemize}
  \item \textbf{Answer Relevancy} (primary metric): Measures how relevant the response is to the question given its context, i.e., the associated persona and KPI.
  \item \textbf{Safety}: Evaluates potential bias in responses. We report the complement of the bias score (Safety $= 1 -$ Bias) for interpretability.
  \item \textbf{Conversation Completeness}: A built-in conversation-level metric that assesses the overall quality of the full question-answer interaction.
  \item \textbf{Custom conversational criteria}: We extend DeepEval with four additional conversation-level metrics to capture dimensions not covered by the default metrics:
  \begin{itemize}
    \item \emph{Focus}: Does the response directly address the specific question?
    \item \emph{Engagement}: Is the language appropriate and engaging for the target persona?
    \item \emph{Helpfulness}: Does the response meaningfully assist the user's analytical task?
    \item \emph{Voice}: Is the response clear and written in active voice?
  \end{itemize}
\end{itemize}

\smallskip\textbf{Success criteria.} Each metric produces a score in $[0, 1]$. A response is considered successful for a given metric if its score meets or exceeds a threshold $\tau$. We define the \emph{success rate} for a metric as the fraction of responses that pass:
\[
  \text{Success Rate} = \frac{|\{i : \text{score}_i \geq \tau\}|}{N}
\]
where $N$ is the total number of evaluated responses. We use the default threshold $\tau = 0.5$ for all metrics in this study to establish a baseline for comparison. We used a default threshold of $0.5$ for all metrics in this study to establish a baseline for comparison. In future, we plan to caliberate thresholds for each metric to better align with user expectations.

\subsection{Evaluation Settings}

We summarize the concrete settings used for question generation and evaluation.

\smallskip\textbf{Database schema.} The experimental database follows a Symitar~\cite{symitar} core banking schema with a tree-like topology. The root table is \texttt{ACCOUNT}, to which eight descendant tables---\texttt{COMMENT}, \texttt{TRACKING}, \texttt{NOTE}, \texttt{SHARE}, \texttt{LOOKUP}, \texttt{LOAN}, \texttt{CARD}, and \texttt{MEMBER}---are joined via left joins on \texttt{ACCOUNT\_NUMBER}. This schema captures a representative range of credit union operations, including member accounts, loan products, card instruments, and account-level annotations.

\smallskip\textbf{Personas and KPIs.} Table~\ref{tab:personas} lists the 13 banking personas and their associated KPIs used in our experiments. Each persona combines two or more roles (e.g., \emph{Risk \& Credit Analytics Manager; CRO; Compliance Officer}) to create realistic variability in analytical focus. The KPIs span risk management (delinquency, non-performing loans), member analytics (tenure, churn, dormancy), product management (portfolio mix, attrition), and fraud detection (limit utilization anomalies).

\begin{table}[t]
\centering
\caption{Banking personas and representative KPIs.}
\label{tab:personas}
\scriptsize
\begin{tabular}{p{3.8cm} p{3.8cm}}
\hline
\textbf{Persona} & \textbf{Representative KPIs} \\
\hline
Risk \& Credit Analytics Mgr; CRO; Compliance & 90+ DPD Rate, Delinquency Ratio, Non-Performing Loan \% \\
Finance / Treasury Mgr; Branch / Regional Mgr & Total Deposit Balance by Geography, Concentration Risk \\
Product Mgr; Member Analytics Lead & Early Closure Rate, Account Churn ($\leq$ 90 days) \\
Finance Mgr; CRO; Product Mgr & Portfolio Mix \%, Credit Card Exposure by Balance \\
Member Analytics Lead; Compliance & Dormant Account Rate, Inactivity Ratio \\
Member Analytics Lead; Executive Leadership & Avg. Account Tenure, Member Lifetime Value \\
Member Analytics Lead; Product Mgr & Active Account Tenure, Retention Duration \\
Product Mgr; Risk Mgr & Monthly Closure Rate, Product Attrition Rate \\
Member Analytics Lead; CRO & Long-Tenure Account \%, Dormant vs Active Ratio \\
Executive Leadership (CEO, COO, Board, CRO) & Product Mix \%, Portfolio Allocation \\
Risk Mgr; Finance Mgr & Composite Health Score (Capital, Asset Quality, Earnings, Liquidity, Growth) \\
Fraud Analyst; Risk Mgr & High-Balance Exposure, Concentration Risk \\
Fraud Analyst; Card Product Mgr & Transfer Limit Exposure, ATM Utilization Anomaly \\
\hline
\end{tabular}
\end{table}

\begin{figure}[t]
\centering
\begin{minipage}{0.95\linewidth}
\scriptsize

\begin{verbatim}
You are a {row['persona']} as {domain} expert.
###
Your goal is to improve the following KPI:
{row['kpi']}.

### Task
Provide actionable {row['difficulty']} difficulty questions
to achieve this goal.

#### {row['difficulty']} difficulty questions must :
{difficulty_meaning}

### Table join
{TABLE_JOINS}

Here are some example questions:
{examples_text}.

### Output Rules (MANDATORY)
- Output JSON only
- Do NOT include explanations, markdown, or extra text
- Each question must be concise, specific, and clearly
tied to the bank domain and the KPI

### Output Schema
{
  "questions": [
    "string",
    "string",
    ...
  ]
}

Generate {n_questions} questions directly in JSON format:
\end{verbatim}
\end{minipage}
\vspace{-0.2cm}
\caption{Prompt template to generate questions.}
\label{fig:question_generation_prompt}
\end{figure}

\begin{figure}[t]
\centering
\begin{minipage}{0.95\linewidth}
\scriptsize

\begin{verbatim}
You are a data analyst specializing in {DOMAIN}.
Answer questions using only commonly available open web knowledge.

Task:
- Answer ONLY the questions listed in the "answerable_question" column.

Rules:
1. Do NOT rephrase or modify the question text.
2. Do NOT use information from any other columns.
3. If a question cannot be answered with general open web data,
return an empty string ("").
4. Limit each answer to 3–5 sentences.
5. Do NOT include citations, sources, opinions, or assumptions.

Output:
Return ONLY valid JSON with no extra text.

Schema:
{
  "answers": [
    {
      "question": "string (exact from answerable_question)",
      "answer": "string (3–5 sentences or empty)"
    }
  ]
}

Generate the JSON response now.
\end{verbatim}
\end{minipage}
\vspace{-0.2cm}
\caption{Prompt template to generate response.}
\vspace{-0.3cm}
\label{fig:response_generation_prompt}
\end{figure}

\smallskip\textbf{Question generation model.} We use GPT-4o-mini with default hyperparameters for synthetic question generation, chosen for its fast inference speed during iterative development. Few-shot examples are randomly sampled from the 15 golden questions, filtered by the target persona and difficulty level. Using this pipeline, we generate approximately 150 synthetic questions spanning all 13 personas and three difficulty levels. We used the prompt template shown in Figure~\ref{fig:question_generation_prompt}.

\smallskip\textbf{Answer generation.} Tursio answers are generated by submitting the synthetic custom questions directly to the Tursio search interface over the Symitar schema. For ChatGPT and Perplexity, the mapped real-world questions (\S2.2) are submitted with a standardized prompt that specifies a 3--5 sentence response length. All answers were collected in December 2025. The prompt template used to generate ChatGPT and Perplexity answers is shown in Figure~\ref{fig:response_generation_prompt}. We generated all the responses in a single turn with no follow-up or clarification allowed, to simulate a realistic search experience where users expect high-quality answers right away.

\smallskip\textbf{Evaluation model.} DeepEval uses GPT-4.1 as the LLM judge for all metrics. The success threshold is set to $\tau = 0.5$ across all metrics. Each response is evaluated with the original question, persona, and KPI provided as context to the judge. We performed single run of evaluation with temperature set to 0 to minimize scoring variability.

\section{Experimental Results}
\label{sec:results}

We now present the evaluation results, starting with an overall comparison of answer relevancy across all three systems, followed by a per-system metric breakdown and a detailed analysis of the score distribution.

\subsection{Overall Answer Relevancy}

\begin{figure}[t]
  \centering
  \includegraphics[width=\columnwidth]{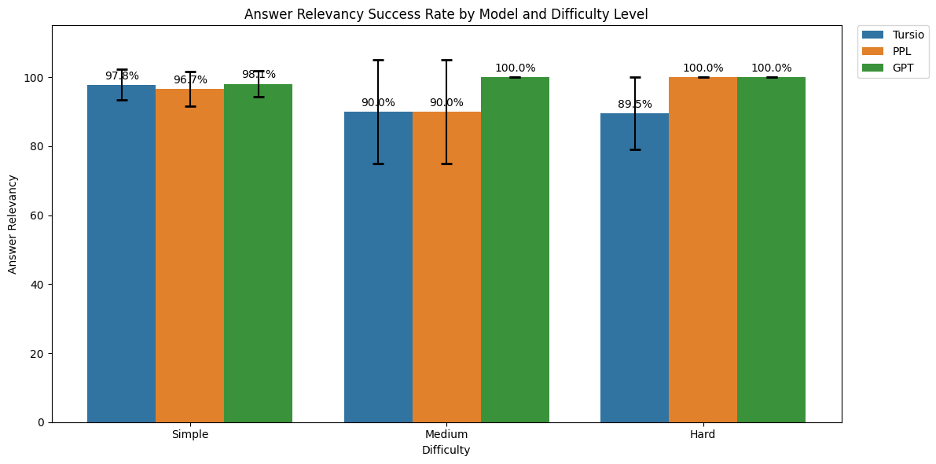}
  \caption{Answer relevancy success rate by system and difficulty level. Tursio achieves relevancy comparable to ChatGPT and Perplexity, with a small gap on medium and hard questions attributable to missing data coverage in the underlying database.}
  \vspace{-0.3cm}
  \Description{Bar chart comparing answer relevancy success rates f
  or Tursio, Perplexity, and ChatGPT across simple, medium, and hard difficulty levels.}
  \label{fig:answer-relevancy}
\end{figure}

Figure~\ref{fig:answer-relevancy} compares the answer relevancy success rate across Tursio, Perplexity (PPL), and ChatGPT (GPT) at each difficulty level. A response is counted as relevant if its answer relevancy score $\geq 0.5$.

On simple questions, all three systems perform comparably: Tursio achieves 97.8\%, Perplexity 96.7\%, and ChatGPT 98.1\%. On medium-difficulty questions, Tursio and Perplexity both achieve 90.0\%, while ChatGPT reaches 100.0\%. On hard questions, Tursio maintains 89.5\%, with Perplexity and ChatGPT both at 100.0\%. The error bars indicate the variability across different samples within each difficulty level.

These results show that Tursio's answer relevancy is competitive with ChatGPT and Perplexity across all difficulty levels, despite operating under a fundamentally different constraint: Tursio answers from a structured database, whereas ChatGPT and Perplexity draw on the open web. The small gap on medium and hard questions is attributable to cases where the underlying database lacks data for certain queries---a limitation inherent to structured data search. Importantly, when Tursio does have the relevant data, its answers are grounded in actual database records, a property that web-based systems cannot guarantee. We analyze the specific metric breakdown in Section~\ref{sec:metric-breakdown} below.

\subsection{Metric Breakdown}
\label{sec:metric-breakdown}

We now examine the full set of evaluation metrics---Safety, Voice, Engagement, Answer Relevancy, Focus, Helpfulness, and Conversation Completeness---for each system across all difficulty levels.

\subsubsection{Tursio}

\begin{figure}[t]
  \centering
  \includegraphics[width=\columnwidth]{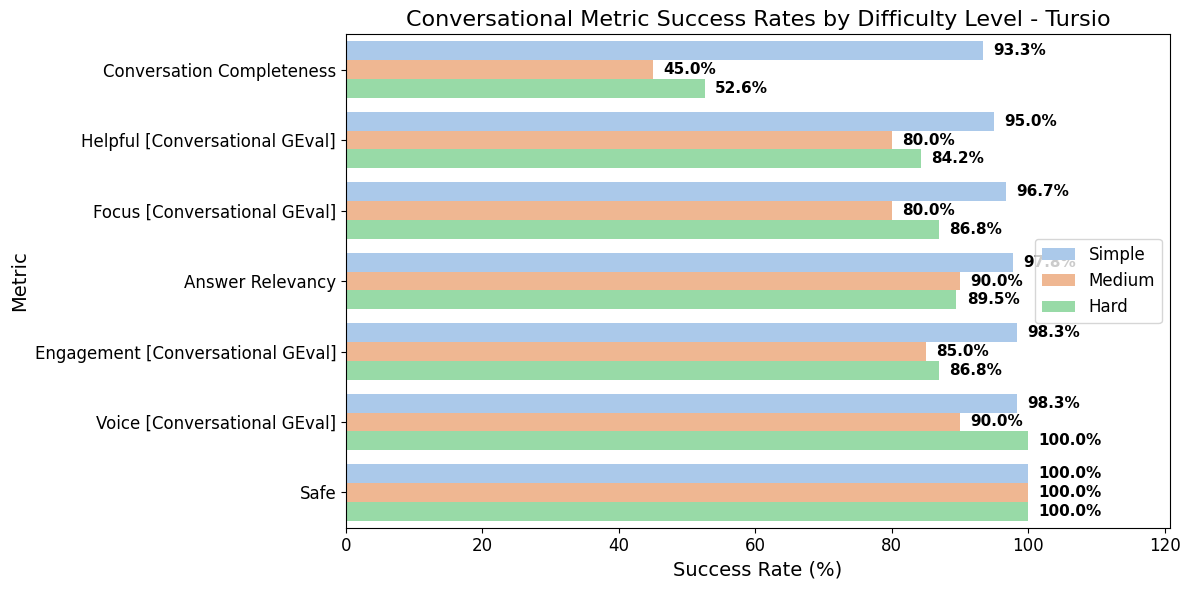}
  \caption{Tursio: per-metric success rates by difficulty level. Most metrics remain above 80\% across all levels, with Conversation Completeness showing the largest gap on medium and hard questions.}
  \Description{Horizontal bar chart showing Tursio's success rates across seven evaluation metrics for simple, medium, and hard difficulty levels.}
  \label{fig:metrics-tursio}
\end{figure}

Figure~\ref{fig:metrics-tursio} shows Tursio's per-metric success rates. On simple questions, Tursio achieves $\geq$93\% for all except the completeness metric, with perfect scores (100\%) on Safety. On medium and hard questions, most metrics remain in the 80--100\% range, with Safety at 100\% and Voice at 90--100\% across all levels. The main exception is Conversation Completeness, which drops to 45.0\% on medium and 52.6\% on hard questions, reflecting cases where the response, while relevant, does not fully address the analytical depth expected by the evaluator. Focus and Helpfulness also show moderate drops on medium questions (both 80.0\%), recovering on hard questions (86.8\% and 84.2\%, respectively).

\smallskip\textbf{Failure analysis.} We identify two root causes for low-scoring responses, illustrated with representative examples:

\emph{(1) Missing data coverage.} When the database lacks the data needed to answer a query, Tursio returns a fallback message (\emph{``no data points were found''}). This is the correct system behavior---Tursio does not hallucinate---but it receives a low relevancy score because the response does not address the question. For example:

\begin{quote}
\small
\textbf{Q:} \emph{What behavioral traits or transaction patterns can predict early warning signals for accounts that transition from 30 DPD to 90 DPD?}\\
\textbf{A:} \emph{Great question but unfortunately no data points were found to generate an answer for you. Consider adjusting your query to change one or more parameters, such as time or other filters.}\\
\textbf{Score:} 0.0 --- The response does not address the question about behavioral traits or transaction patterns.
\end{quote}

This pattern accounts for the majority of failures on medium and hard questions, where queries often require data columns (e.g., geographic region, promotional offer metadata) that are absent from the Symitar schema.

\emph{(2) Semantic mismatch.} In some cases, Tursio retrieves relevant data but answers a slightly different question than the one asked. For example:

\begin{quote}
\small
\textbf{Q:} \emph{What is the distribution of credit card balances across different age groups of members?}\\
\textbf{A:} \emph{The dataset shows significant variability in credit card limits across the member groups\ldots the average credit card limit ranges from \$2,109 (Group 40) to \$7,294 (Group 66)\ldots}\\
\textbf{Score:} 0.0 --- The response discusses credit card \emph{limits}, not \emph{balances} as requested.
\end{quote}

These semantic mismatches are less frequent than missing data failures but represent an area for improvement in Tursio's query interpretation.

\subsubsection{Perplexity}

\begin{figure}[t]
  \centering
  \includegraphics[width=\columnwidth]{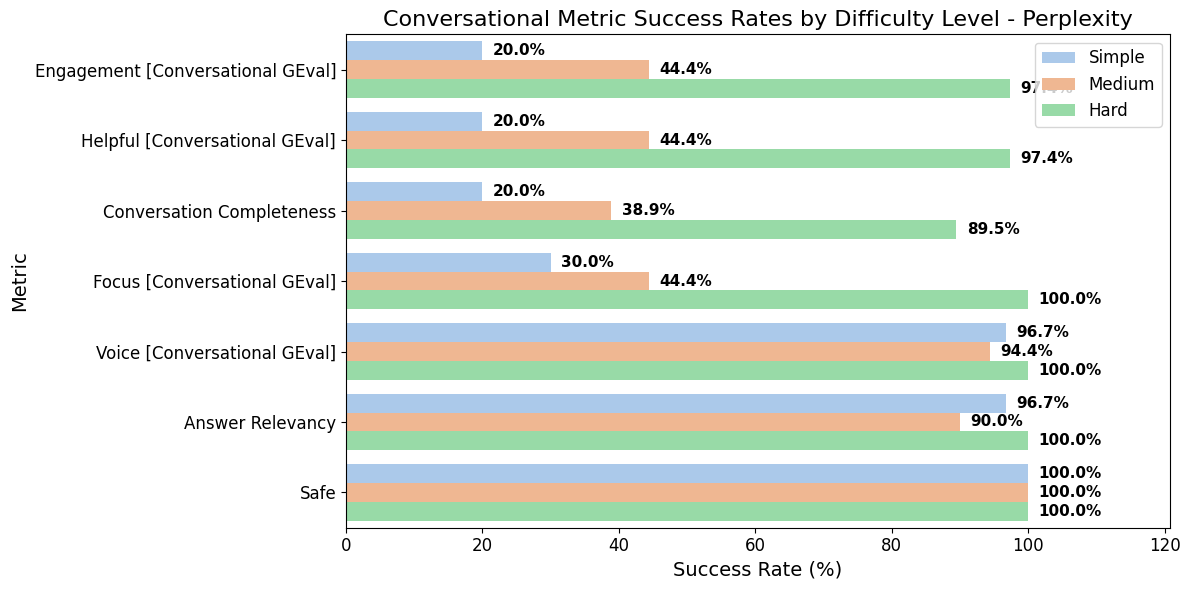}
  \caption{Perplexity: per-metric success rates by difficulty level. Voice and Answer Relevancy are consistently high, while conversational metrics are lower on simple questions due to the constrained response length.}
  \Description{Horizontal bar chart showing Perplexity's success rates across seven evaluation metrics for simple, medium, and hard difficulty levels.}
  \label{fig:metrics-perplexity}
\end{figure}

Figure~\ref{fig:metrics-perplexity} shows Perplexity's per-metric success rates. Perplexity achieves perfect scores (100\%) on Safety across all difficulty levels, and scores highly on Voice (96.7--100\%) and Answer Relevancy (90.0--100\%). However, the conversational metrics reveal a notable pattern: on simple questions, Completeness (20.0\%), Helpfulness (20.0\%), Focus (30.0\%), and Engagement (20.0\%) are all significantly lower than on hard questions, where these metrics reach 89.5--97.4\%. This inversion---where hard questions score better than simple ones---likely reflects the standardized 3--5 sentence response length: for simple, factual questions, concise responses are penalized by the completeness evaluator for not elaborating, whereas complex questions naturally benefit from the same response length providing richer analytical content.

\begin{figure}[t]
  \centering
  \includegraphics[width=\columnwidth]{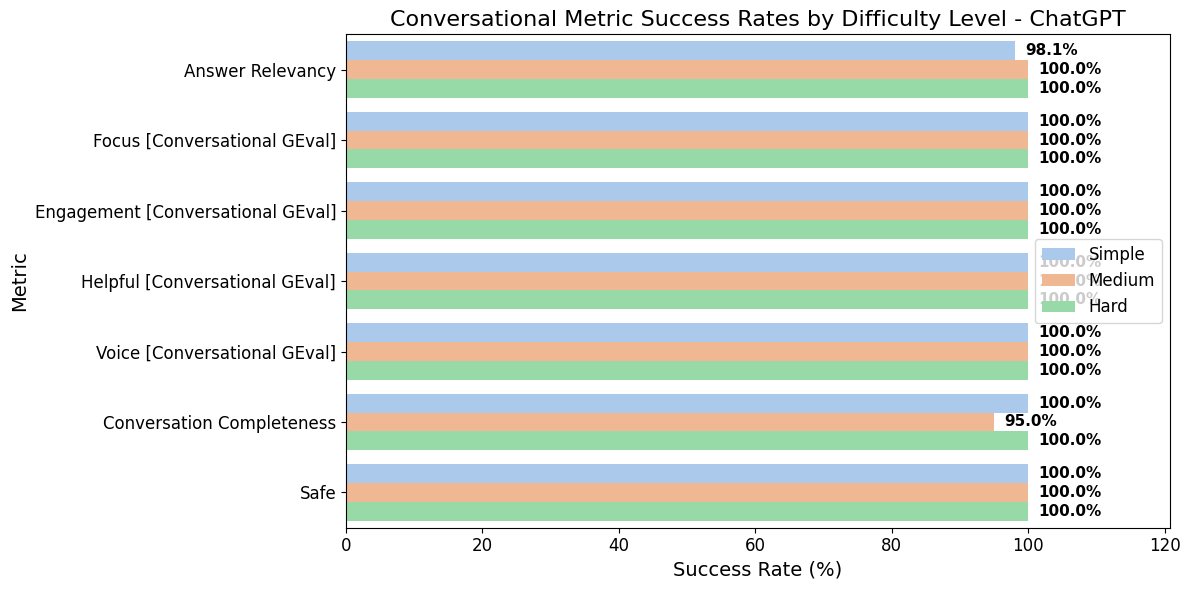}
  \caption{ChatGPT: per-metric success rates by difficulty level. ChatGPT achieves near-perfect scores across all metrics, establishing an upper-bound baseline for search quality.}
  \Description{Horizontal bar chart showing ChatGPT's success rates across seven evaluation metrics for simple, medium, and hard difficulty levels.}
  \label{fig:metrics-chatgpt}
\end{figure}

\subsubsection{ChatGPT}

Figure~\ref{fig:metrics-chatgpt} shows ChatGPT's per-metric success rates. ChatGPT achieves near-perfect performance across all metrics and difficulty levels, with 100\% success on Safety, Voice, Helpfulness, Focus, and Engagement at every difficulty level. Conversation Completeness reaches 100\% on simple and hard questions, with a minor dip to 95.0\% on medium. Answer Relevancy is 98.1\% on simple questions and 100\% on medium and hard. These results establish ChatGPT as a strong upper-bound baseline: operating on open-domain data with well-tuned conversational responses, it consistently satisfies all evaluation criteria.

\subsection{Score Distribution}

While the success rates in the previous section indicate whether responses pass a threshold, they do not reveal how scores are distributed. A system with 90\% success rate could have most scores clustered near 1.0 (high confidence) or spread across the passing range (borderline). In this section, we examine the metric score histograms for each system to understand the shape and concentration of scores across difficulty levels.

\subsubsection{Tursio}

\begin{figure}[t]
  \centering
  \includegraphics[width=\columnwidth]{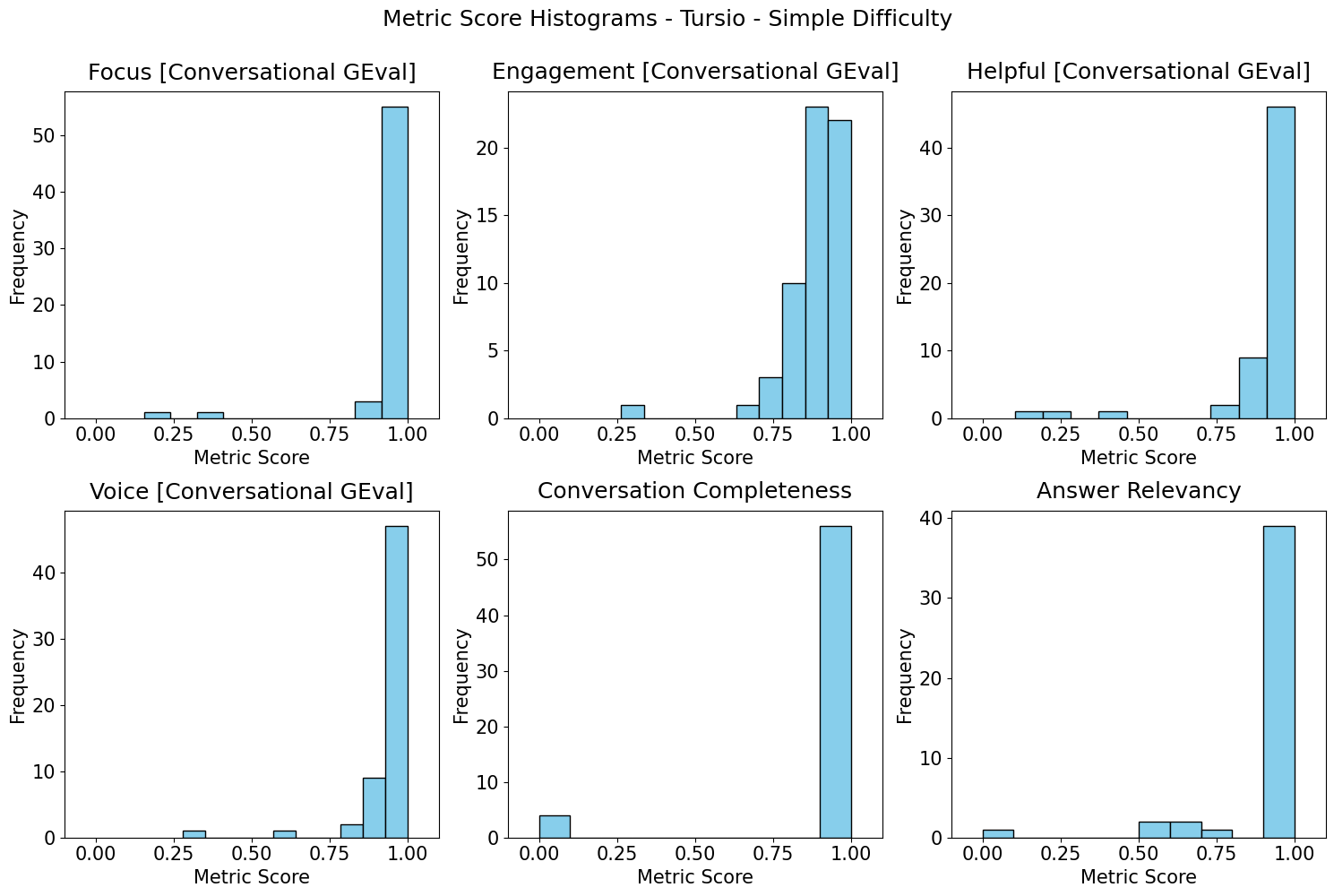}
  \caption{Tursio score distributions on simple questions. Scores are sharply concentrated near 1.0 across all metrics.}
  \Description{Histograms of Tursio metric scores for simple questions showing right-skewed distributions concentrated near 1.0.}
  \label{fig:hist-tursio-simple}
\end{figure}

\begin{figure}[t]
  \centering
  \includegraphics[width=\columnwidth]{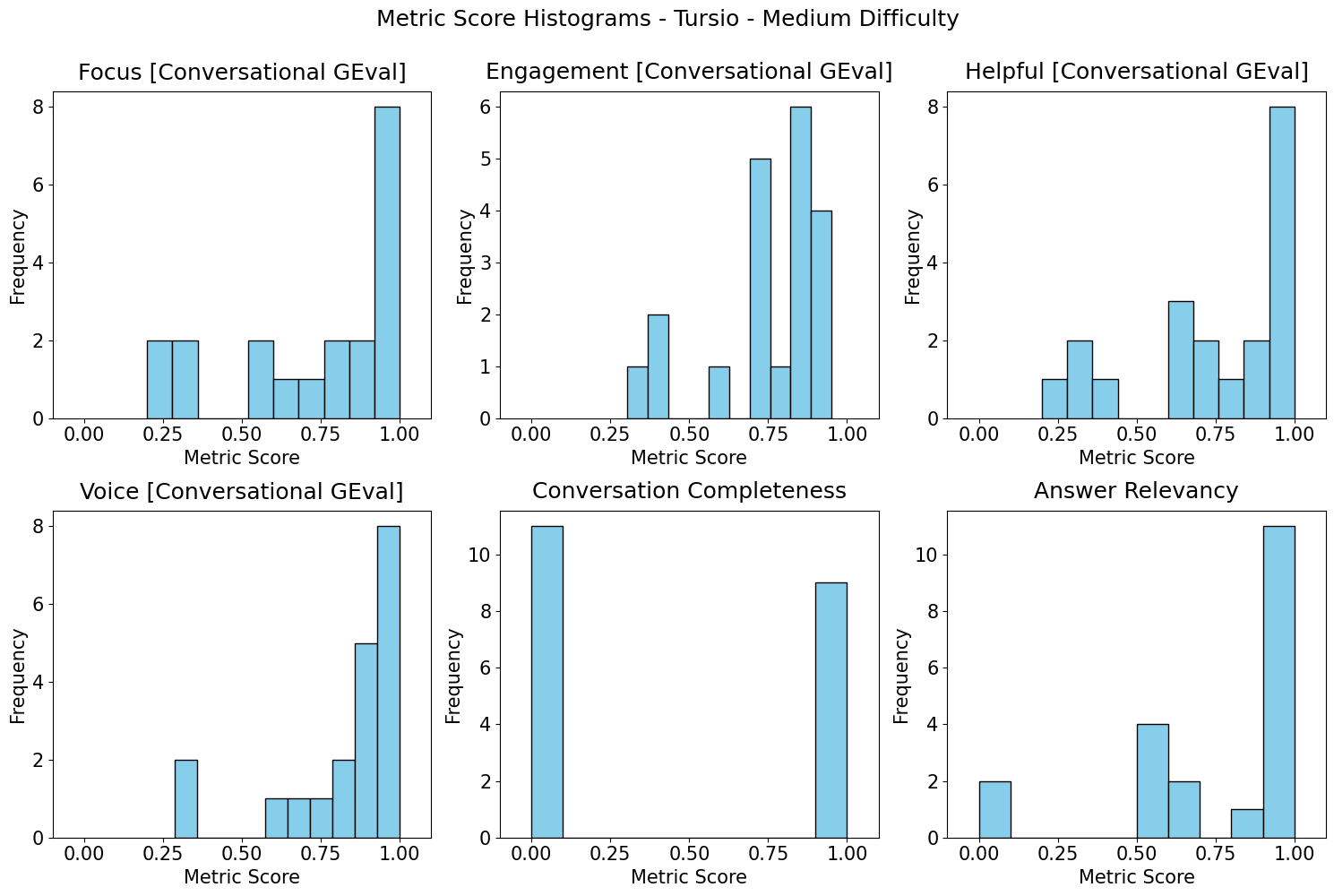}
  \caption{Tursio score distributions on medium questions. Conversation Completeness shows a bimodal pattern due to data gaps; other metrics show moderate spread.}
  \Description{Histograms of Tursio metric scores for medium questions showing wider distributions and bimodal completeness.}
  \label{fig:hist-tursio-medium}
\end{figure}

\begin{figure}[t]
  \centering
  \includegraphics[width=\columnwidth]{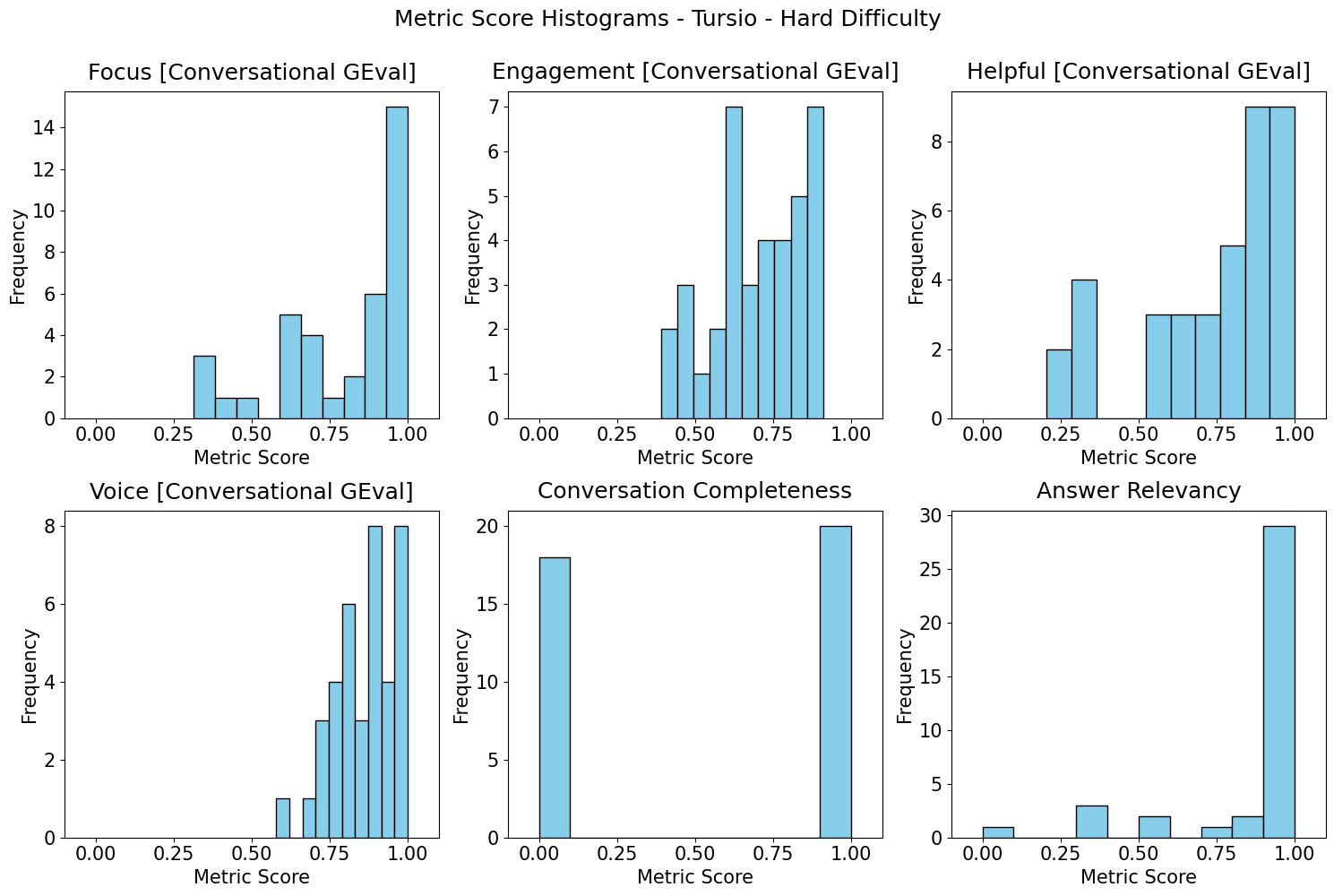}
  \caption{Tursio score distributions on hard questions. The bimodal completeness pattern persists; other metrics remain predominantly high with longer tails.}
  \Description{Histograms of Tursio metric scores for hard questions showing bimodal completeness and broader engagement distributions.}
  \label{fig:hist-tursio-hard}
\end{figure}

Figures~\ref{fig:hist-tursio-simple}--\ref{fig:hist-tursio-hard} show Tursio's score distributions across difficulty levels.
On simple questions (Figure~\ref{fig:hist-tursio-simple}), scores are sharply concentrated near 1.0 for all metrics. Answer Relevancy, Voice, Focus, and Helpfulness show strong right-skewed distributions with the vast majority of responses scoring above 0.9. Conversation Completeness is similarly concentrated at the high end. The only minor spread appears in Engagement, where a small number of responses score in the 0.6--0.8 range.

On medium questions (Figure~\ref{fig:hist-tursio-medium}), the distributions become notably wider. Conversation Completeness exhibits a bimodal pattern: a cluster near 0.0 (responses where the database lacked sufficient data to provide a complete answer) and a cluster near 1.0. Answer Relevancy remains right-skewed but shows a few responses near 0.0, corresponding to the missing-data failures discussed in Section~\ref{sec:metric-breakdown}. Focus and Helpfulness show moderate spread across the 0.2--1.0 range, reflecting the partial-answer cases where Tursio addresses some but not all aspects of a multi-part question.

\begin{figure}[t]
  \centering
  \includegraphics[width=\columnwidth]{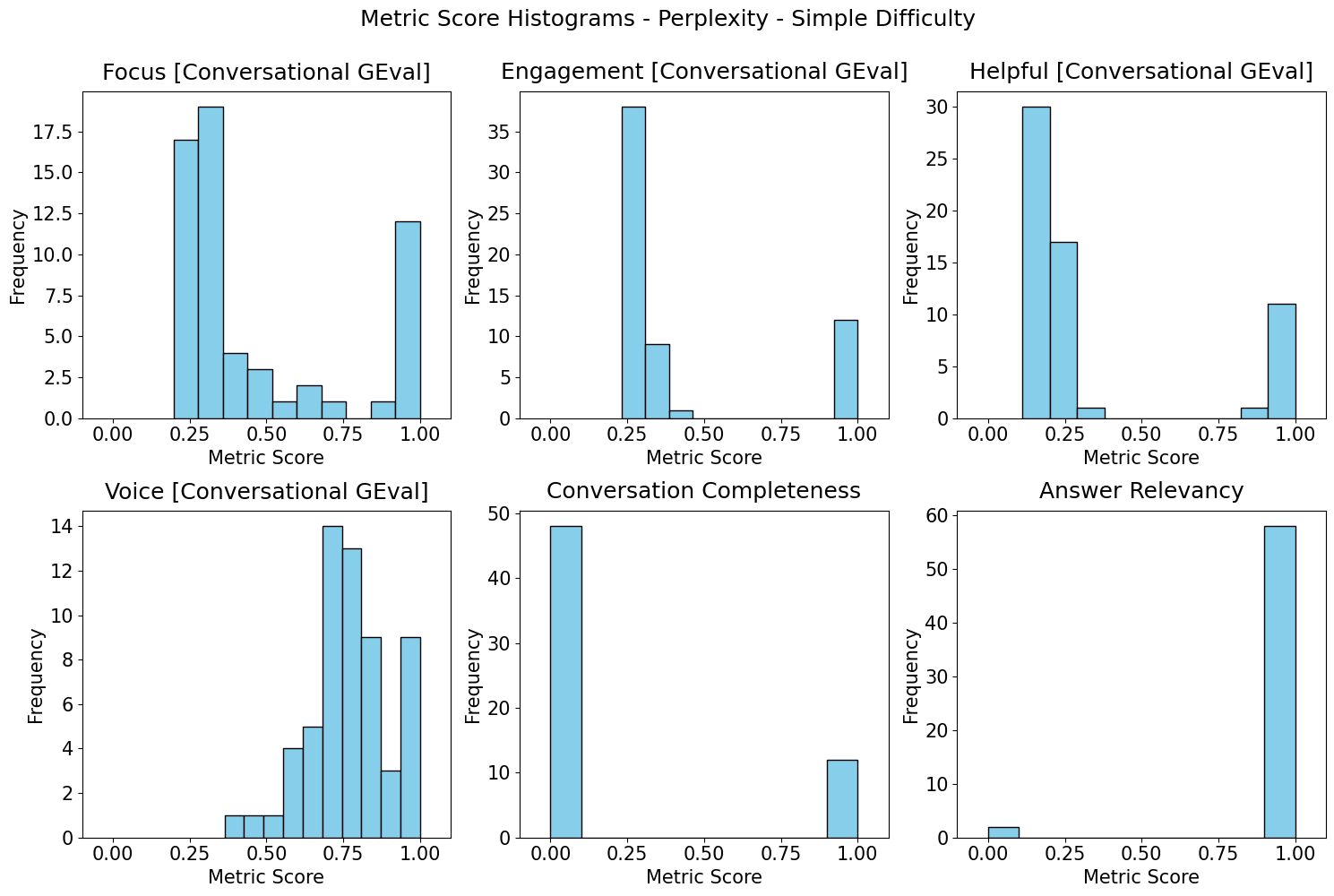}
  \caption{Perplexity score distributions on simple questions. Answer Relevancy clusters near 1.0, but conversational metrics (Focus, Engagement, Helpfulness, Completeness) are concentrated at the low end due to brief responses.}
  \Description{Histograms of Perplexity metric scores for simple questions showing high relevancy but low conversational metric scores.}
  \label{fig:hist-ppl-simple}
\end{figure}

On hard questions (Figure~\ref{fig:hist-tursio-hard}), the bimodal pattern in Conversation Completeness persists, with roughly equal mass at the low and high ends. Answer Relevancy remains predominantly right-skewed with most scores near 1.0, but a visible tail extends to 0.0--0.3. Engagement and Focus show broader distributions centered around 0.7--0.9, indicating more variability in conversational quality for complex analytical questions. Voice remains consistently high, clustering in the 0.8--1.0 range.

\subsubsection{Perplexity}

\begin{figure}[t]
  \centering
  \includegraphics[width=\columnwidth]{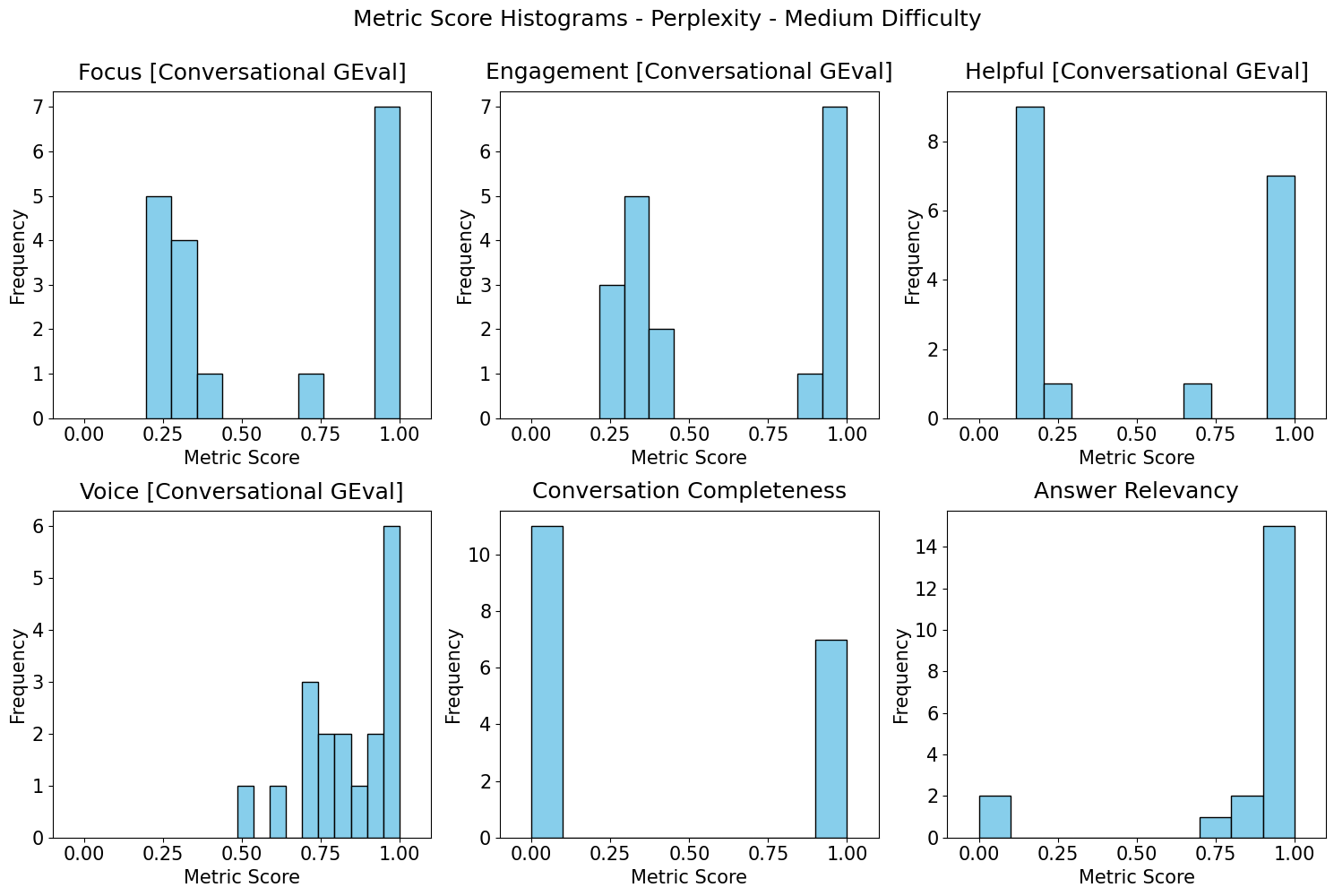}
  \caption{Perplexity score distributions on medium questions. Conversational metrics exhibit bimodal distributions, splitting between low-scoring and high-scoring clusters, while Answer Relevancy remains concentrated near 1.0.}
  \Description{Histograms of Perplexity metric scores for medium questions showing bimodal distributions for conversational metrics and right-skewed Answer Relevancy.}
  \label{fig:hist-ppl-medium}
\end{figure}

\begin{figure}[t]
  \centering
  \includegraphics[width=\columnwidth]{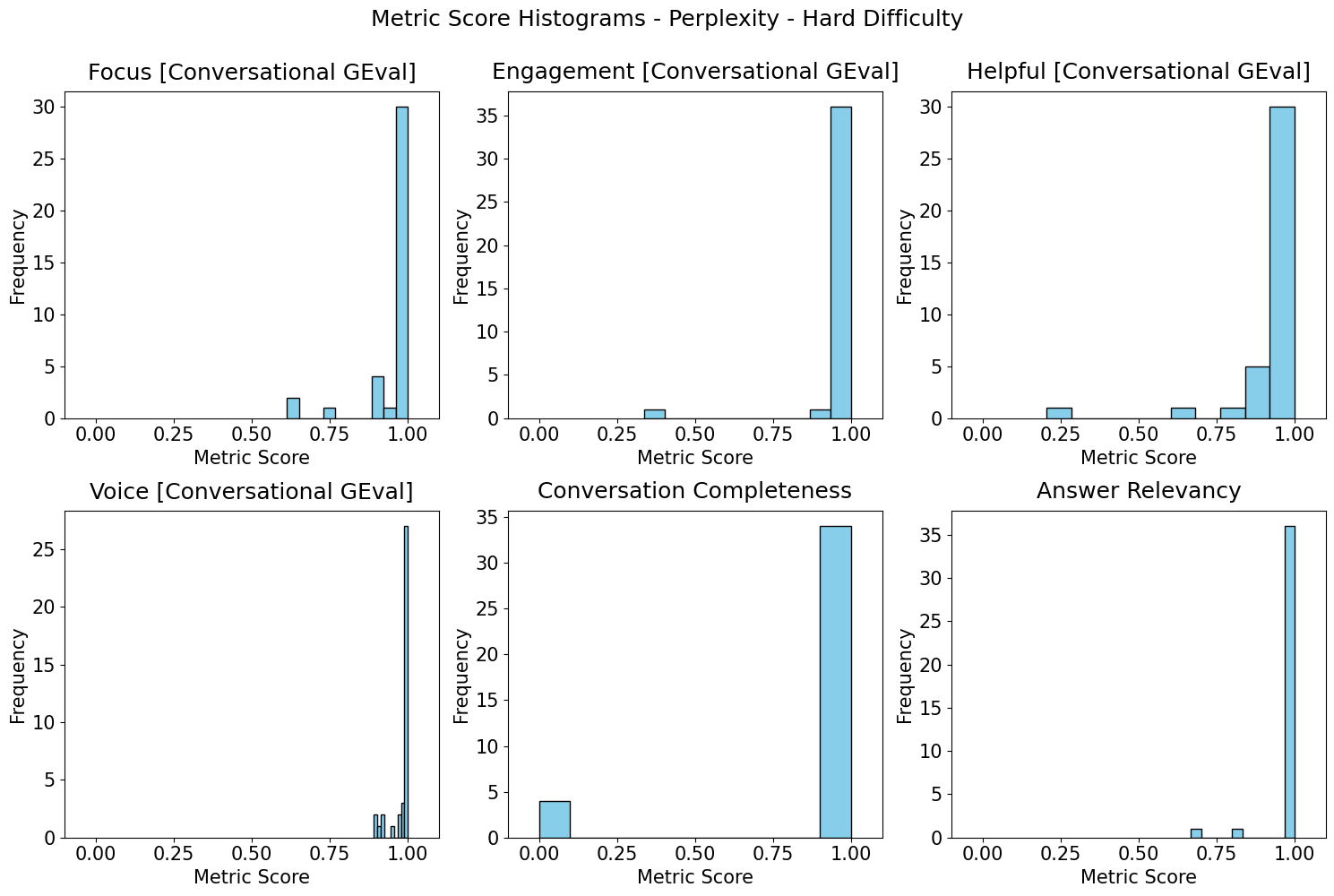}
  \caption{Perplexity score distributions on hard questions. All metrics shift to the high end, as complex questions elicit more detailed and analytically rich responses.}
  \Description{Histograms of Perplexity metric scores for hard questions showing all distributions concentrated near 1.0.}
  \label{fig:hist-ppl-hard}
\end{figure}

Figures~\ref{fig:hist-ppl-simple}--\ref{fig:hist-ppl-hard} show Perplexity's score distributions.
On simple questions (Figure~\ref{fig:hist-ppl-simple}), a striking divergence appears between metrics. Answer Relevancy is sharply concentrated near 1.0, and Voice scores are distributed in the 0.6--1.0 range. However, the conversational metrics---Focus, Engagement, Helpfulness, and Completeness---are concentrated at the \emph{low} end, with most scores falling in the 0.1--0.4 range. This pattern reflects Perplexity's concise response style: for simple factual questions, short answers score well on relevancy but are penalized by conversational metrics that expect more elaboration, context, and analytical depth.

On medium questions (Figure~\ref{fig:hist-ppl-medium}), the distributions reveal a transitional pattern. Answer Relevancy remains strongly right-skewed with most scores near 1.0, confirming that Perplexity's factual accuracy holds at this difficulty level. Voice scores distribute across 0.5--1.0, clustering near the high end. However, the conversational metrics---Focus, Engagement, Helpfulness, and Completeness---exhibit pronounced bimodal distributions: each splits between a low-scoring cluster (0.1--0.3) and a high-scoring cluster near 1.0. Helpfulness is particularly striking, with roughly 9 responses near 0.1 and 5 near 1.0, with almost nothing in between. Completeness shows a similar pattern, with a dominant cluster near 0.0 and a smaller cluster near 1.0. This bimodality suggests that medium questions divide into two groups: those that map well to publicly available information (yielding detailed, high-scoring responses) and those that remain too domain-specific for comprehensive web-based answers.

On hard questions (Figure~\ref{fig:hist-ppl-hard}), Perplexity's distributions improve markedly. All metrics concentrate near 1.0, with Focus, Engagement, Helpfulness, and Completeness all shifting to the high end. Complex analytical questions naturally elicit more detailed responses from Perplexity, which satisfy the conversational evaluation criteria. A small number of Completeness scores near 0.0 remain, corresponding to cases where the response scope was insufficient.

\subsubsection{ChatGPT}

\begin{figure}[t]
  \centering
  \includegraphics[width=\columnwidth]{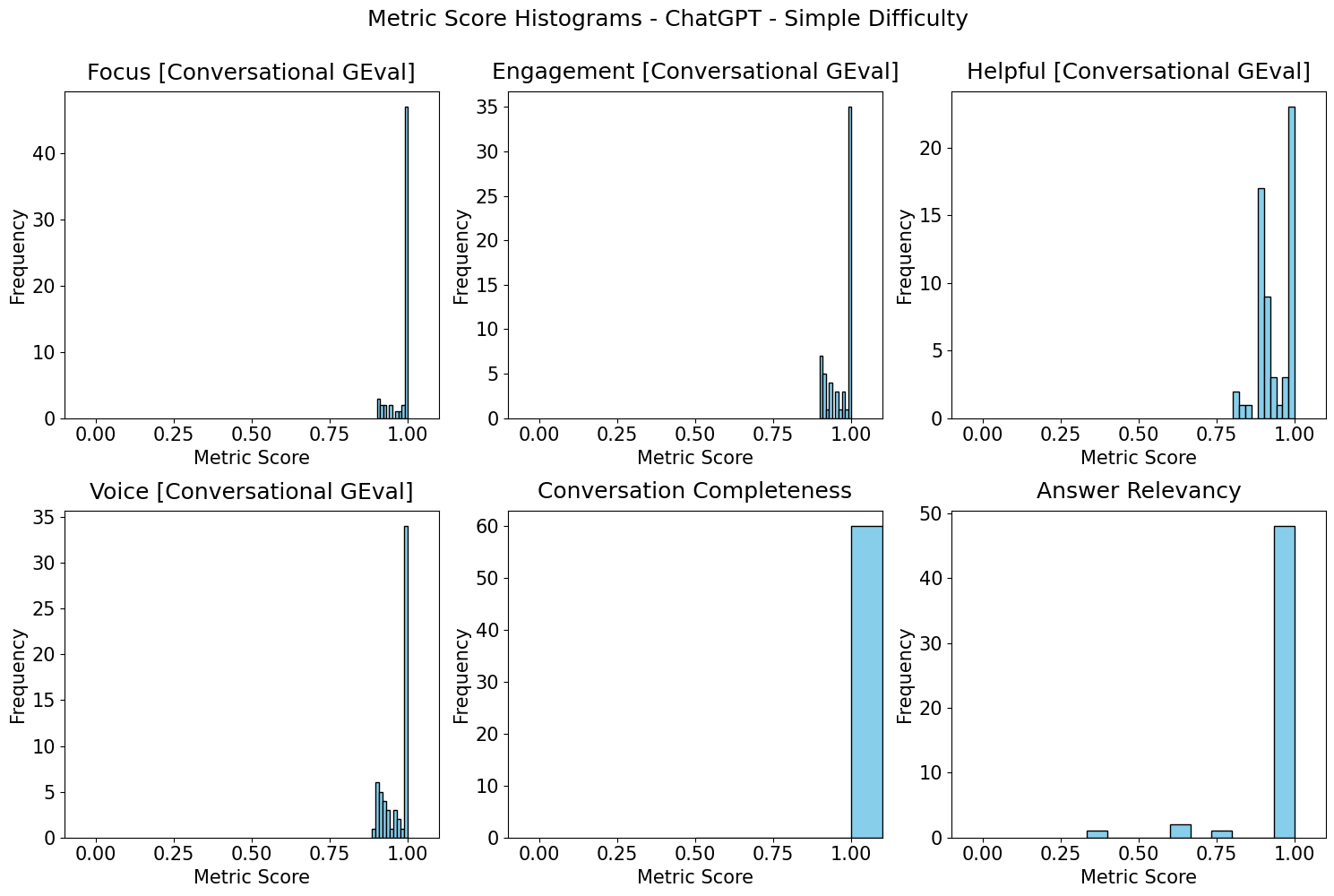}
  \caption{ChatGPT score distributions on simple questions. All metrics are sharply concentrated near 1.0, with Conversation Completeness at a perfect 1.0.}
  \Description{Histograms of ChatGPT metric scores for simple questions showing tight distributions near 1.0.}
  \label{fig:hist-gpt-simple}
\end{figure}

\begin{figure}[t]
  \centering
  \includegraphics[width=\columnwidth]{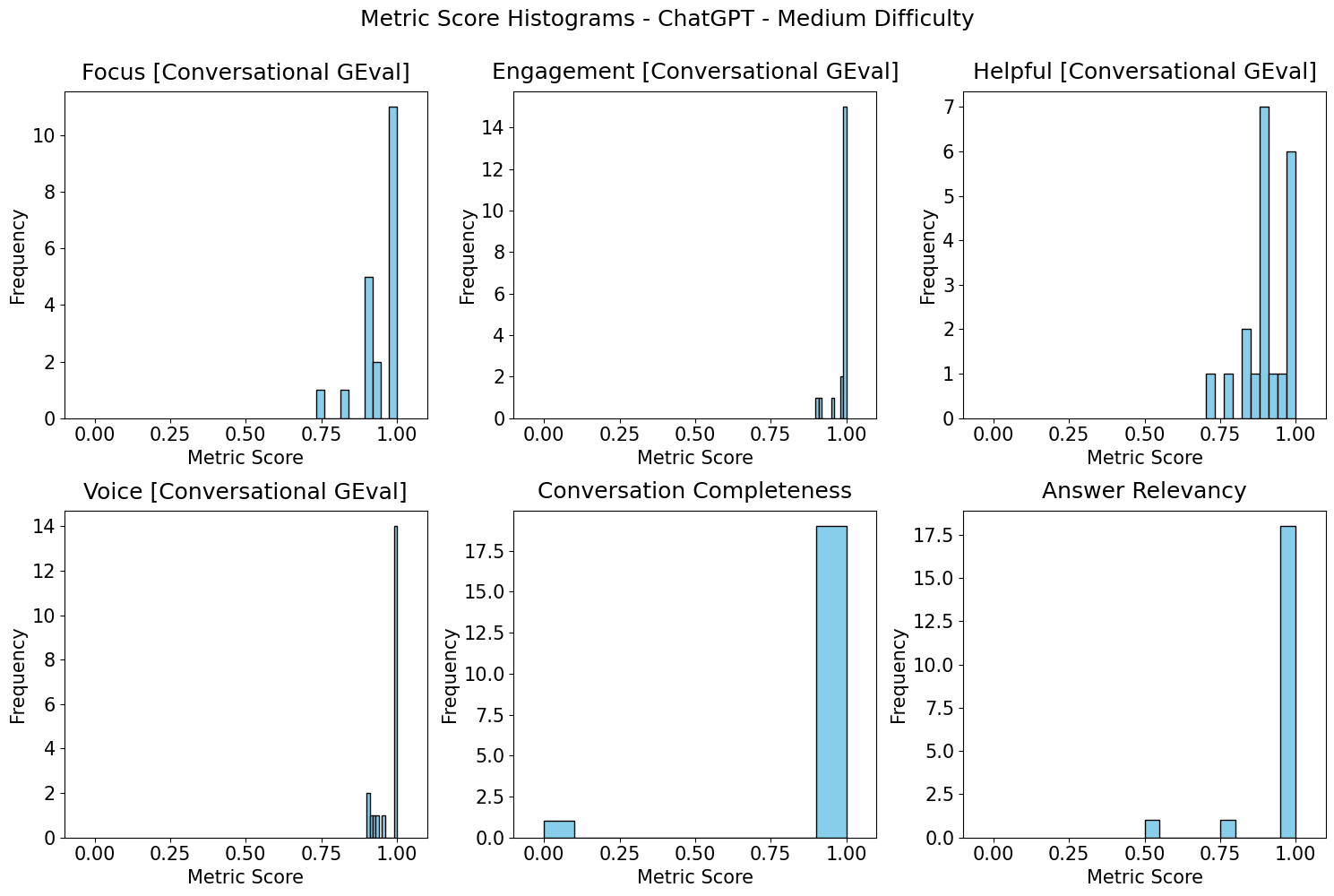}
  \caption{ChatGPT score distributions on medium questions. All metrics remain tightly concentrated near 1.0, with a single Conversation Completeness outlier near 0.0 accounting for the 95\% success rate.}
  \Description{Histograms of ChatGPT metric scores for medium questions showing tight distributions near 1.0 with one completeness outlier.}
  \label{fig:hist-gpt-medium}
\end{figure}

\begin{figure}[t]
  \centering
  \includegraphics[width=\columnwidth]{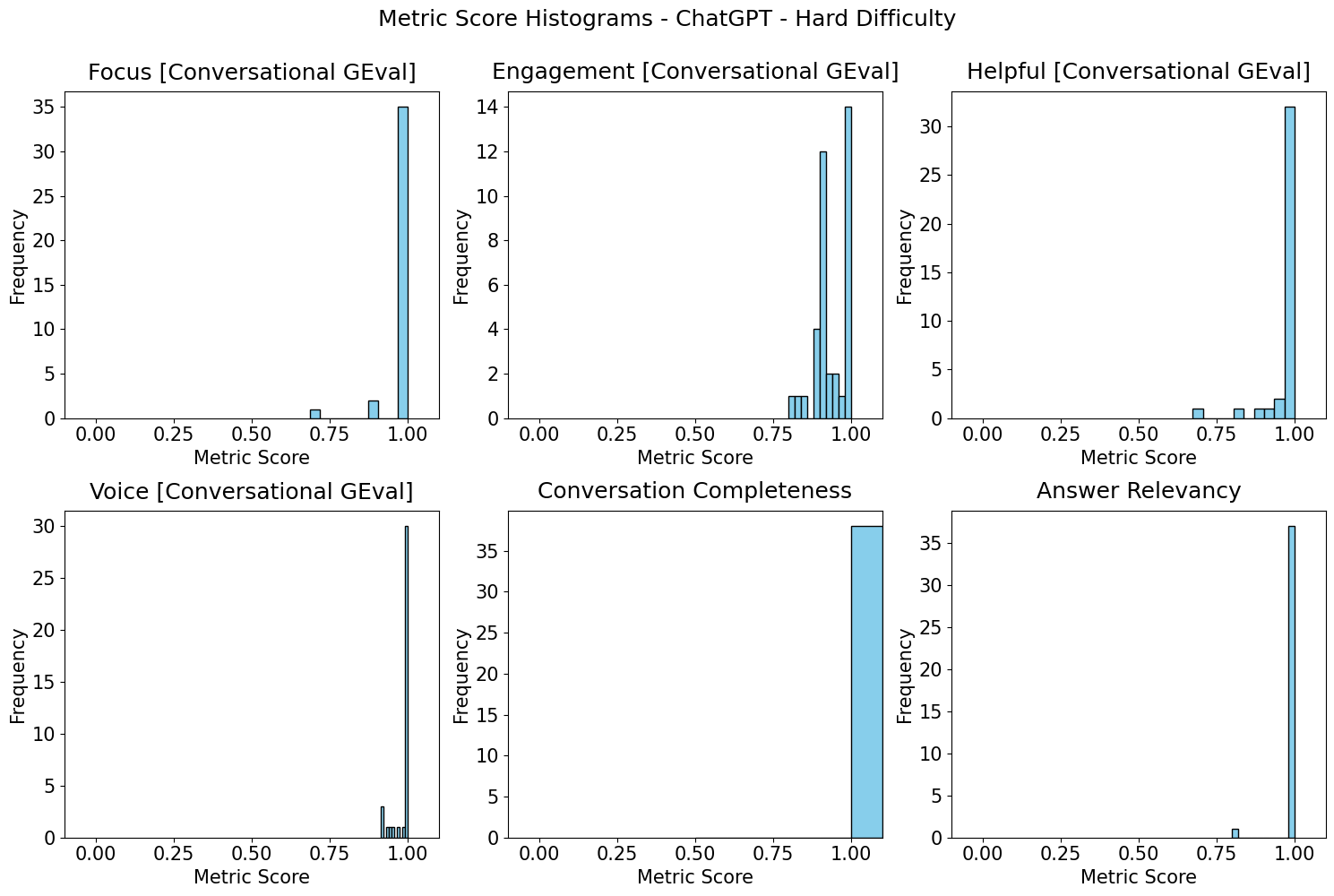}
  \caption{ChatGPT score distributions on hard questions. Engagement shows the widest spread; all other metrics remain concentrated near 1.0.}
  \Description{Histograms of ChatGPT metric scores for hard questions showing tight distributions with minor engagement spread.}
  \label{fig:hist-gpt-hard}
\end{figure}

Figures~\ref{fig:hist-gpt-simple}--\ref{fig:hist-gpt-hard} show ChatGPT's score distributions.
On simple questions (Figure~\ref{fig:hist-gpt-simple}), ChatGPT's distributions are sharply right-skewed across all metrics, with the vast majority of scores at or near 1.0. Conversation Completeness is perfectly concentrated at 1.0. Answer Relevancy, Focus, and Helpfulness show a tight cluster at 1.0 with only a handful of responses scoring slightly lower (0.8--1.0). Engagement shows the most spread, with scores distributed across 0.9--1.0, indicating minor variability in conversational tone.

On medium questions (Figure~\ref{fig:hist-gpt-medium}), the pattern is nearly identical. Answer Relevancy and Voice remain tightly concentrated near 1.0, with over 18 and 14 responses at the peak, respectively. Engagement clusters at 1.0 with minimal spread. Focus and Helpfulness show slight variability in the 0.8--0.9 range but remain strongly right-skewed. The only notable deviation is a single Conversation Completeness response near 0.0, which accounts for the minor dip to 95.0\% success rate on this metric---the sole imperfection in an otherwise near-perfect profile.

On hard questions (Figure~\ref{fig:hist-gpt-hard}), ChatGPT maintains its strong performance. Engagement shows the widest spread, with scores distributed across 0.8--1.0, reflecting the challenge of maintaining conversational quality on complex analytical questions. All other metrics remain tightly clustered at 1.0. These consistently concentrated distributions confirm ChatGPT's role as a strong upper-bound baseline in our evaluation.

\subsection{Statistical Significance}

To determine whether the observed differences in answer relevancy are statistically meaningful, we apply pairwise chi-square tests comparing Tursio against each baseline (ChatGPT and Perplexity) at each difficulty level. For each comparison, we construct a $2 \times 2$ contingency table of success/failure counts and compute the chi-square statistic with Yates' correction.

\begin{figure}[t]
  \centering
  \includegraphics[width=\columnwidth]{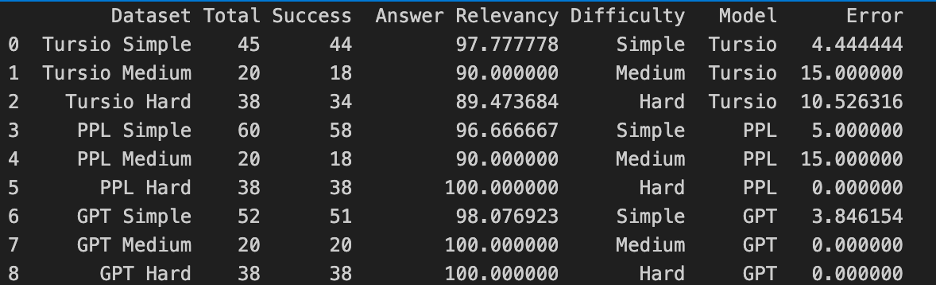}
  \caption{Summary of answer relevancy evaluation results. For each system and difficulty level: total samples evaluated, number of successful responses (score $\geq 0.5$), answer relevancy rate (\%), and error margin.}
  \Description{Table showing sample sizes, success counts, answer relevancy percentages, and error margins for Tursio, Perplexity, and ChatGPT across simple, medium, and hard difficulty levels.}
  \label{fig:stat-sig}
\end{figure}

Figure~\ref{fig:stat-sig} summarizes the sample sizes, success counts, answer relevancy rates, and error margins for each system-difficulty combination. The sample sizes vary slightly across systems (e.g., 45 for Tursio Simple vs.\ 60 for PPL Simple) due to occasional API timeouts during answer generation with DeepEval.

Table~\ref{tab:chi-square} reports the pairwise chi-square test results. Across all six comparisons, no statistically significant difference is found at the $\alpha = 0.05$ level. On simple questions, both comparisons yield $p = 1.0$, confirming that the three systems are virtually indistinguishable. On medium questions, Tursio vs.\ ChatGPT yields $p = 0.47$ and Tursio vs.\ Perplexity yields $p = 1.0$. The smallest p-values occur on hard questions ($p = 0.12$ for both comparisons), where ChatGPT and Perplexity achieve 100\% relevancy compared to Tursio's 89.5\%---but even this gap does not reach statistical significance given the sample sizes.

These results confirm that \textbf{Tursio's answer relevancy is statistically comparable to ChatGPT and Perplexity} across all difficulty levels, despite the fundamental difference in data access: Tursio queries a structured database while the baselines draw on the open web.

\begin{table}[t]
\centering
\caption{Pairwise chi-square tests for answer relevancy success rates. No comparison reaches statistical significance ($\alpha = 0.05$).}
\label{tab:chi-square}
\small
\begin{tabular}{llcc}
\hline
\textbf{Difficulty} & \textbf{Comparison} & \textbf{$p$-value} & \textbf{Significant?} \\
\hline
Simple & Tursio vs.\ GPT & 1.0000 & No \\
Simple & Tursio vs.\ PPL & 1.0000 & No \\
Medium & Tursio vs.\ GPT & 0.4682 & No \\
Medium & Tursio vs.\ PPL & 1.0000 & No \\
Hard   & Tursio vs.\ GPT & 0.1233 & No \\
Hard   & Tursio vs.\ PPL & 0.1233 & No \\
\hline
\end{tabular}
\end{table}

\section{Performance Analysis}
\label{sec:analysis}

Having presented the quantitative results, we now synthesize the findings into a deeper analysis of each system's strengths, weaknesses, and the factors that drive their performance. We examine how data availability, response style, and query complexity interact to produce the observed patterns.

\subsection{Tursio}

\smallskip\textbf{Answer relevancy and contextual understanding.} Tursio's answer relevancy is high on simple questions (97.8\%) and degrades modestly on medium (90.0\%) and hard (89.5\%) questions. The failures fall into two categories identified in Section~\ref{sec:metric-breakdown}: missing data coverage and semantic mismatch. For simple questions, isolated failures occur when Tursio misinterprets the query context. For example, responding about loan records when the question asks about account comments. On medium and hard questions, the dominant failure mode shifts to data unavailability, where the underlying database simply does not contain the columns or records needed to answer the query. This distinction is important: the gap is primarily a \emph{data coverage} issue rather than a \emph{comprehension} issue, and can be addressed by enriching the database or improving the fallback responses that inform users better.

\smallskip\textbf{Completeness as the key bottleneck.} Conversation Completeness is Tursio's weakest metric, dropping from 93.3\% on simple questions to 45.0\% on medium and 52.6\% on hard questions. This degradation is strongly correlated with query complexity: medium and hard questions often request multi-faceted analyses (e.g., breakdowns across multiple dimensions, trend comparisons over time), and when the database provides only partial data, Tursio can address some but not all aspects of the question. For instance, a question requesting five account details may receive a response covering only one, yielding a partial but technically relevant answer. This results in high Answer Relevancy but low Completeness---a pattern that suggests Tursio prioritizes precision over recall in its responses.

\smallskip\textbf{Conversational quality.} Tursio's conversational metrics: Engagement (85.0--98.3\%), Helpfulness (80.0--95.0\%), and Focus (80.0--96.7\%)---remain strong on simple questions and degrade moderately on medium and hard questions. The drop is largely driven by the same data availability issues: when Tursio lacks sufficient data, its responses become less helpful and less focused. Notably, Voice remains consistently high (90.0--100\%), indicating that Tursio's response style and clarity are not the limiting factor.

\smallskip\textbf{Consistency.} The score distribution analysis (Section~3.3) reveals that Tursio exhibits higher variance than the baselines, particularly on medium and hard questions. While Tursio's upper-bound scores match those of ChatGPT and Perplexity, the longer tails in its distributions indicate less predictable performance. This variance is driven by the binary nature of data availability: queries that hit well-populated parts of the schema receive excellent scores, while those that encounter data gaps receive poor scores.

\subsection{Perplexity}

\smallskip\textbf{The open-data bifurcation.} Perplexity exhibits a distinctive performance pattern: high Answer Relevancy across all difficulty levels (90.0--100\%) coexists with low conversational metrics on simple and medium questions. On simple questions, Engagement (20.0\%), Helpfulness (20.0\%), Completeness (20.0\%), and Focus (30.0\%) are all critically low despite a 96.7\% answer relevancy rate. This bifurcation is interesting and it reverses on hard questions, where all metrics rise above 89\%.

This pattern has a clear explanation: simple banking questions (e.g., ``What percentage of accounts have made a comment in the last year?'') are domain-specific and difficult to answer comprehensively using public web data. The mapped open-domain equivalents may lack the specificity needed for a detailed, engaging response. Hard questions, by contrast, tend to be more analytical and research-oriented (e.g., trend analyses, risk comparisons), which align better with the kind of content available on the open web.

\smallskip\textbf{Response format constraints.} Perplexity consistently achieves high Voice scores (94.4--100\%), confirming that its response quality and clarity are strong. However, the standardized 3--5 sentence response limit creates a systematic tension with the Completeness metric. For questions requiring multi-part answers, this brevity constraint prevents comprehensive coverage, even when Perplexity correctly identifies the relevant information. However, this is an evaluation artifact rather than a fundamental limitation: longer responses would likely improve Completeness without degrading other metrics.

\subsection{ChatGPT}

\smallskip\textbf{Consistent high performance.} ChatGPT achieves near-perfect scores across all metrics and difficulty levels, with most metrics at 100\% and only minor dips in Answer Relevancy on simple questions (98.1\%) and Conversation Completeness on medium questions (95.0\%). This consistency makes ChatGPT the strongest baseline in our evaluation and reflects the advantages of operating on open-domain data with a highly tuned conversational model.

\smallskip\textbf{Score concentration.} The score distributions (Section~3.3) confirm that ChatGPT's high success rates are not borderline: scores are tightly concentrated near 1.0, with minimal spread. The only notable variance appears in Engagement on hard questions, where scores distribute across 0.8--1.0 rather than clustering at 1.0. This suggests that even ChatGPT faces minor challenges in maintaining optimal conversational tone on complex analytical questions, though the impact on success rates is negligible.

\smallskip\textbf{The open-data advantage.} ChatGPT's strong performance must be interpreted in context: it answers open-domain equivalents of the banking questions, drawing on its extensive pretraining data. It does not face the data availability constraints that limit Tursio (structured database) or the data dependency issues that affect Perplexity (web search). This makes ChatGPT an informative upper bound but obviously not a direct substitute for Tursio's structured data search capability.

\subsection{Summary}

The analysis reveals three distinct performance profiles shaped by each system's data access model:

\begin{itemize}
  \item \textbf{Tursio} is bounded by database completeness. When the data exists, Tursio delivers precise, grounded answers competitive with the baselines. When data is missing, performance degrades---primarily in Completeness rather than Relevancy, indicating that Tursio correctly avoids hallucination at the cost of partial answers.

  \item \textbf{Perplexity} is bounded by open-web data alignment. Its performance varies based on how well the query maps to publicly available information, creating an inverted difficulty curve where hard, analytical questions outperform simple, domain-specific ones on conversational metrics.

  \item \textbf{ChatGPT} benefits from broad pretraining data and strong conversational tuning, achieving consistent high performance across all conditions. It serves as an upper-bound baseline, though it cannot provide the database-grounded answers that Tursio delivers.
\end{itemize}

The key finding is that despite these different constraints, Tursio achieves \emph{statistically comparable} answer relevancy to both baselines (Section~3.4), demonstrating that structured database search can match the quality of open-domain systems on enterprise data.

\section{Future Directions}
\label{sec:future}

Our evaluation reveals several areas for improvement, both in the evaluation methodology itself and in the systems being evaluated. We organize these into three directions.

\smallskip\textbf{Richer evaluation methodology.} The current evaluation has three key limitations that future iterations should address. First, the single-turn QA format does not capture multi-turn conversational patterns that are common in real-world search sessions, where users refine queries, ask follow-up questions, and request clarifications. Extending the evaluation to multi-turn interactions (3--5 exchanges) would better assess context retention, clarification handling, and conversational coherence. Second, the standardized 3--5 sentence response limit, while ensuring comparability, systematically penalizes systems on the Completeness metric for questions that require detailed, multi-part answers. Future evaluations should experiment with unconstrained response lengths to disentangle format constraints from genuine completeness gaps. Third, the current LLM-as-judge approach (GPT-4.1) introduces its own biases. Incorporating human-aligned reward models from benchmarks such as RewardBench~\cite{RewardBench} would reduce evaluator subjectivity and improve correlation with human judgments.

\smallskip\textbf{Improving data coverage and question quality.} The analysis in Section~4 identifies database completeness as Tursio's primary bottleneck, with Conversation Completeness dropping to 45.0\% on medium questions due to data gaps. Addressing this requires systematic coverage audits across query patterns, data enrichment for identified gaps (particularly account-level details and temporal data), and improved fallback responses that provide more informative explanations when data is unavailable. On the question generation side, the golden question set should be expanded to 3--5 examples per persona-difficulty combination to ensure balanced coverage and reduce evaluator uncertainty. The question mapping step for open-domain benchmarking should also be refined, as the current mapping occasionally produces questions that are difficult for web-based systems to answer comprehensively---particularly for simple, domain-specific queries where Perplexity's conversational metrics drop to 20--30\%.

\smallskip\textbf{Scaling and automation.} The current evaluation pipeline involves manual steps in answer collection that limit scalability and reproducibility. Future work should automate the end-to-end pipeline---from question generation through answer collection to metric computation---enabling continuous evaluation as the underlying systems and databases evolve. This includes automated response extraction via APIs, version-controlled datasets for reproducibility, and monitoring dashboards for tracking quality trends over time.

\section{Related Work}

We survey related work across four areas: natural language interfaces to databases, text-to-SQL benchmarks, AI-powered search engines, and evaluation methodologies for LLM-based systems.

\subsection{Natural Language Interfaces to Databases}

The vision of querying databases in natural language dates back to the 1970s, with early systems like LUNAR~\cite{woods1972lunar} and Codd's RENDEZVOUS~\cite{codd1974rendezvous}. PRECISE~\cite{precise} formalized the problem by reducing semantic interpretation to a maximum-bipartite-matching problem, guaranteeing correct SQL generation for a broad class of ``semantically tractable'' questions. NaLIR~\cite{nalir} extended the approach to support complex SQL features (aggregation, nesting, various join types) through interactive disambiguation with the user.

The advent of deep learning transformed NLIDB research into the NL2SQL paradigm, where the focus shifted to training models that translate natural language to SQL. Recent approaches leverage LLMs with decomposition strategies: DIN-SQL~\cite{dinsql} breaks text-to-SQL into sub-tasks (schema linking, query classification, SQL generation) and achieves 85.3\% execution accuracy on Spider, while DAIL-SQL uses efficient few-shot prompting to reach 86.6\%.

However, as Floratou et al.~\cite{nl2sql_solved_not} argue in their CIDR 2024 paper ``NL2SQL is a Solved Problem... Not!'', enterprise-grade NL2SQL remains far from resolved. Key challenges include schema complexity, semantic ambiguity, and the mismatch between academic benchmarks and production workloads. Production systems from Databricks (Genie)~\cite{databricks-genie-knowledge-store}, Snowflake (Cortex Analyst)~\cite{snowflake-views-semantic-overview}, and Google (Gemini)~\cite{google-gemini-database-understanding} address this gap through semantic layers, knowledge stores, and agentic reasoning---but they focus on SQL generation rather than evaluating the quality of the final answer returned to the business user.

Looking beyond single databases, Eckmann and Binnig~\cite{qcp-db} propose a vision for \emph{query-by-collaboration}, where autonomous data agents---each wrapping a distinct data source---collaboratively answer natural language queries without traditional data integration. Their Query Collaboration Protocol (QCP) allows agents to advertise capabilities, exchange partial results, and synthesize answers across heterogeneous databases, achieving accuracy comparable to centralized systems on a modified BIRD benchmark. While their work addresses the data integration bottleneck for multi-source querying, it does not evaluate the quality of the final natural language answer returned to the user.

Our focus in this paper is evaluate the \emph{end-to-end search experience}, measuring whether the final natural language answer is relevant, complete, and useful to the business user, rather than whether the intermediate SQL is correct.

\subsection{Text-to-SQL Benchmarks}

Text-to-SQL benchmarks have evolved from simple single-table datasets to complex, cross-domain evaluations. WikiSQL~\cite{wikisql} introduced large-scale evaluation with 80K question-SQL pairs but is limited to simple single-table queries. Spider~\cite{spider_benchmark} advanced the field significantly with 10K questions across 200 databases requiring cross-domain generalization. SParC~\cite{sparc} and CoSQL~\cite{cosql} extended Spider to multi-turn conversational settings, evaluating context retention across question sequences. KaggleDBQA~\cite{kaggledbqa} introduced evaluation over real Kaggle databases with domain-specific data types and unrestricted questions.

As discussed in Section~2.1, BIRD~\cite{bird_bench} is the most widely used benchmark, with 12,751 pairs across 95 databases, and BEAVER~\cite{beaver_benchmark} provides a more realistic benchmark where state-of-the-art LLMs achieve far lower execution accuracy.

A common limitation across all these benchmarks is their evaluation metric: execution accuracy (whether predicted and gold SQL produce identical result sets). This metric does not assess whether the system produces a correct, complete, and understandable natural language answer---the quality dimension that matters most to end users and that our evaluation framework is designed to measure. Our evaluation closes this gap by assessing final answer quality rather than intermediate SQL correctness.

\subsection{AI-Powered Search Engines}

A new generation of AI-powered search engines has emerged that combine LLMs with web retrieval to provide synthesized answers rather than ranked document lists. Perplexity AI generates cited, synthesized answers from web sources and reports 93.9\% accuracy on SimpleQA~\cite{simpleqa} via its Deep Research feature. Exa~\cite{exa_ai} trains end-to-end neural architectures for web search and has developed domain-specific evaluation benchmarks for people and company search. Tavily~\cite{tavily} provides a search API purpose-built for AI agents and RAG workflows, achieving 93.3\% accuracy on SimpleQA by feeding retrieved content to GPT-4.1. You.com~\cite{you_com} targets enterprise AI search and evaluates across accuracy, freshness, latency, and cost dimensions using SimpleQA, FreshQA, and MS MARCO benchmarks. And so on.

These systems evaluate search quality primarily through factual accuracy on open-domain QA benchmarks. Our work addresses a complementary setting: search over \emph{structured enterprise databases}, where the challenge is not web retrieval but mapping high-level business questions to the correct database structures and returning grounded, precise answers. Tursio bridges these two worlds by delivering a comparable search experience over structured data.

\subsection{Evaluation Methodologies}

Automated evaluation of LLM-generated text has advanced rapidly along two fronts. The \emph{LLM-as-judge} paradigm, introduced by Zheng et al.~\cite{mtbench} with MT-Bench and Chatbot Arena, uses strong LLMs to evaluate chat assistants, achieving over 80\% agreement with human preferences. AlpacaEval~\cite{alpacaeval} refines this approach with length-controlled comparisons to mitigate verbosity bias. G-Eval~\cite{geval} applies chain-of-thought reasoning for NLG evaluation, achieving strong correlation with human judgments on summarization tasks.

For retrieval-augmented systems, RAGAS~\cite{ragas} introduces reference-free metrics that separately evaluate retrieval (context precision, context recall) and generation (faithfulness, answer relevancy) components. TruLens provides a complementary ``RAG Triad'' of groundedness, answer relevance, and contextual relevance metrics.

Our evaluation builds on this foundation---using DeepEval~\cite{deepeval}, which implements the G-Eval methodology---but adapts it to the structured database search setting. Unlike standard RAG evaluation, where the retrieval corpus is a collection of text documents, our setting involves structured tables with schema semantics, join relationships, and domain-specific conventions. We also extend beyond answer relevancy to include conversational metrics (Completeness, Focus, Engagement, Helpfulness, Voice) that capture the full search experience. To our knowledge, this is the first application of LLM-as-judge evaluation to structured database search.

\subsection{Table Question Answering}

Table QA benchmarks evaluate question answering over semi-structured tabular data. WikiTableQuestions~\cite{wikitablequestions} introduced large-scale evaluation with 22K question-answer pairs over Wikipedia tables requiring aggregation, comparison, and sorting. HotpotQA~\cite{hotpotqa} introduced 113K Wikipedia-based question-answer pairs that require \emph{multi-hop reasoning}---finding and combining evidence across multiple documents---with sentence-level supporting-fact annotations that enable explainability evaluation. HybridQA~\cite{hybridqa} extended multi-hop reasoning to the hybrid setting of tables and linked text passages. OTT-QA~\cite{ottqa} further scaled to the open-domain setting with 45K questions over 400K candidate tables.

While table QA shares the goal of answering questions over structured data, it operates at a fundamentally different scale and complexity than database search. Table QA benchmarks typically involve individual tables or small collections with extractive answers, whereas enterprise database search requires navigating normalized multi-table schemas with complex joins, aggregations, and domain-specific semantics---as demonstrated by the 30--55$\times$ SQL-to-question token-length ratios in production workloads (Figure~\ref{fig:token-ratio}). Our evaluation targets this more complex setting, where answers must be synthesized across multiple joined tables rather than extracted from a single one.



\section{Conclusion}

We set out to answer a simple question: can a structured database search system deliver a search experience comparable to ChatGPT? Our evaluation shows that the answer is yes---Tursio achieves answer relevancy statistically indistinguishable from both ChatGPT and Perplexity across simple, medium, and hard enterprise queries, despite the fundamental constraint of answering from a structured database rather than the open web.

This result is significant because it demonstrates that the search quality gap between enterprise database systems and consumer AI search engines is narrower than commonly assumed. The remaining gaps are not in comprehension or response quality but in data coverage: when the database has the data, Tursio's answers are precise, grounded, and competitive; when it does not, performance degrades gracefully through fallbacks rather than hallucination.

More broadly, our work highlights the need for evaluation frameworks that go beyond SQL correctness to measure what actually matters to business users: whether the answer is relevant, complete, and useful. As natural language interfaces to databases become ubiquitous, the community needs rigorous, end-to-end evaluation methodologies for understanding user experience in structured data search. 
We hope this work provides a step toward that direction.


\balance
\bibliographystyle{ACM-Reference-Format}
\bibliography{references}

@online{sun-times-ai-misinformation,
  author = {Chicago Sun-Times},
  title = {Syndicated Content: Sunday Print Sun-Times AI Misinformation},
  year = {2025},
  url = {https://chicago.suntimes.com/news/2025/05/20/syndicated-content-sunday-print-sun-times-ai-misinformation},
  lastaccessed = {July 19, 2025},
}

@online{bbc-ai-mistakes,
  author = {BBC},
  title = {AI mistakes: People are getting paid to fix them},
  year = {2025},
  url = {https://www.bbc.com/news/articles/cyvm1dyp9v2o},
  lastaccessed = {July 19, 2025},
}

@inproceedings{bird_bench,
 author = {Li, Jinyang and others},
 booktitle = {Advances in Neural Information Processing Systems},
 title = {Can LLM Already Serve as A Database Interface? A BIg Bench for Large-Scale Database Grounded Text-to-SQLs},
 year = {2023}
}

@misc{beaver_benchmark,
      title={BEAVER: An Enterprise Benchmark for Text-to-SQL}, 
      author={Peter Baile Chen and others},
      year={2025},
      archivePrefix={arXiv},
      url={https://arxiv.org/abs/2409.02038}, 
}

@online{databricks-genie-knowledge-store,
  author = {Databricks},
  title = {Databricks Genie Knowledge Store},
  year = {2025},
  url = {https://docs.databricks.com/aws/en/genie/knowledge-store},
  lastaccessed = {October 7, 2025},
}

@online{snowflake-views-semantic-overview,
  author = {Snowflake},
  title = {Overview of Semantic Views},
  year = {2025},
  url = {https://docs.snowflake.com/en/user-guide/views-semantic/overview},
  lastaccessed = {November 15, 2025},
}

@online{google-gemini-database-understanding,
  author = {Google Cloud},
  title = {A new top score: Advancing Text-to-SQL on the BIRD benchmark},
  year = {2025},
  lastaccessed = {November 14, 2025},
  url = {https://cloud.google.com/blog/products/databases/how-to-get-gemini-to-deeply-understand-your-database}
}

@article{ProfilingRelationalData15,
  author = {Abedjan, Ziawasch and Golab, Lukasz and Naumann, Felix},
  title = {Profiling relational data: a survey},
  year = {2015},
  publisher = {Springer-Verlag},
  volume = {24},
  number = {4},
  journal = {VLDB Journal},
  pages = {557–581},
  numpages = {25}
}

@article{DatabasesSearchableDeepContext15,
  author = {Alekh Jindal and others},
  title = {Making Databases Searchable with Deep Context},
  year={2026},
  eprint={2602.08320},
  archivePrefix={arXiv},
  primaryClass={cs.DB},
  url={https://arxiv.org/abs/2602.08320}
}

@online{symitar,
  author = {Jack Henry},
  title = {Symitar Core Banking Platform},
  year = {2025},
  url = {https://www.jackhenry.com/what-we-offer/core/symitar},
  lastaccessed = {February 16, 2026},
}

@online{deepeval,
  author = {Confident AI},
  title = {DeepEval: The Open-Source LLM Evaluation Framework},
  year = {2024},
  url = {https://github.com/confident-ai/deepeval},
  lastaccessed = {February 16, 2026},
}

@inproceedings{geval,
  author = {Yang Liu and others},
  title = {G-Eval: NLG Evaluation using GPT-4 with Better Human Alignment},
  booktitle = {Proceedings of the 2023 Conference on Empirical Methods in Natural Language Processing},
  year = {2023},
}

@inproceedings{spider_benchmark,
  author = {Yu, Tao and others},
  title = {Spider: A Large-Scale Human-Labeled Dataset for Complex and Cross-Domain Semantic Parsing and Text-to-SQL Task},
  booktitle = {EMNLP},
  year = {2018},
}

@article{wikisql,
  author = {Zhong, Victor and Xiong, Caiming and Socher, Richard},
  title = {Seq2SQL: Generating Structured Queries from Natural Language using Reinforcement Learning},
  journal = {arXiv:1709.00103},
  year = {2017},
}

@inproceedings{sparc,
  author = {Yu, Tao and others},
  title = {SParC: Cross-Domain Semantic Parsing in Context},
  booktitle = {ACL},
  year = {2019},
}

@inproceedings{cosql,
  author = {Yu, Tao and others},
  title = {CoSQL: A Conversational Text-to-SQL Challenge Towards Cross-Domain Natural Language Interfaces to Databases},
  booktitle = {EMNLP},
  year = {2019},
}

@inproceedings{kaggledbqa,
  author = {Lee, Chia-Hsuan and Polozov, Oleksandr and Richardson, Matthew},
  title = {KaggleDBQA: Realistic Evaluation of Text-to-SQL Parsers},
  booktitle = {ACL},
  year = {2021},
}

@inproceedings{nalir,
  author = {Li, Fei and Jagadish, H. V.},
  title = {Constructing an Interactive Natural Language Interface for Relational Databases},
  journal = {PVLDB},
  volume = {8},
  number = {1},
  pages = {73--84},
  year = {2014},
}

@inproceedings{precise,
  author = {Popescu, Ana-Maria and Etzioni, Oren and Kautz, Henry},
  title = {Towards a Theory of Natural Language Interfaces to Databases},
  booktitle = {Proceedings of the 8th International Conference on Intelligent User Interfaces},
  year = {2003},
  pages = {149--157},
}

@inproceedings{nl2sql_solved_not,
  author = {Floratou, Avrilia and others},
  title = {NL2SQL is a Solved Problem... Not!},
  booktitle = {CIDR},
  year = {2024},
}

@inproceedings{dinsql,
  author = {Pourreza, Mohammadreza and Rafiei, Davood},
  title = {DIN-SQL: Decomposed In-Context Learning of Text-to-SQL with Self-Correction},
  booktitle = {NeurIPS},
  year = {2023},
}

@inproceedings{ragas,
  author = {Es, Shahul and James, Jithin and Espinosa-Anke, Luis and Schockaert, Steven},
  title = {RAGAS: Automated Evaluation of Retrieval Augmented Generation},
  booktitle = {EACL},
  year = {2024},
}

@inproceedings{mtbench,
  author = {Zheng, Lianmin and others},
  title = {Judging LLM-as-a-Judge with MT-Bench and Chatbot Arena},
  booktitle = {NeurIPS},
  year = {2023},
}

@inproceedings{alpacaeval,
  author = {Dubois, Yann and Galambosi, Bal\'{a}zs and Liang, Percy and Burns, Tatsunori B.},
  title = {Length-Controlled AlpacaEval: A Simple Way to Debias Automatic Evaluators},
  booktitle = {COLM},
  year = {2024},
}

@inproceedings{wikitablequestions,
  author = {Pasupat, Panupong and Liang, Percy},
  title = {Compositional Semantic Parsing on Semi-Structured Tables},
  booktitle = {ACL},
  year = {2015},
}

@inproceedings{hybridqa,
  author = {Chen, Wenhu and others},
  title = {HybridQA: A Dataset of Multi-Hop Question Answering over Tabular and Textual Data},
  booktitle = {Findings of EMNLP},
  year = {2020},
}

@inproceedings{ottqa,
  author = {Chen, Wenhu and Chang, Ming-Wei and Schlinger, Eva and Wang, William Yang and Cohen, William W.},
  title = {Open Question Answering over Tables and Text},
  booktitle = {ICLR},
  year = {2021},
}

@inproceedings{hotpotqa,
  author = {Yang, Zhilin and others},
  title = {HotpotQA: A Dataset for Diverse, Explainable Multi-hop Question Answering},
  booktitle = {EMNLP},
  year = {2018},
}

@inproceedings{qcp-db,
  author = {Eckmann, Timo and Binnig, Carsten},
  title = {A Vision for Autonomous Data Agent Collaboration: From Query-by-Integration to Query-by-Collaboration},
  booktitle = {CIDR},
  year = {2026},
}

@online{exa_ai,
  author = {Exa},
  title = {Exa: The Search Engine for AIs},
  year = {2025},
  url = {https://exa.ai},
  lastaccessed = {February 17, 2026},
}

@online{tavily,
  author = {Tavily},
  title = {Tavily: Search API for AI Agents},
  year = {2025},
  url = {https://tavily.com},
  lastaccessed = {February 17, 2026},
}

@online{you_com,
  author = {You.com},
  title = {You.com AI Search API},
  year = {2025},
  url = {https://you.com},
  lastaccessed = {February 17, 2026},
}

@inproceedings{RewardBench,
  author = {Nathan Lambert and others},
  title = {RewardBench: Evaluating Reward Models for Language Modeling},
  booktitle = {Proceedings of the 62nd Annual Meeting of the Association for Computational Linguistics},
  year = {2024},
}

@misc{simpleqa,
      title={SimpleQA: Measuring Short-form Factuality in Large Language Models},
      author={Jason Wei and others},
      year={2024},
      archivePrefix={arXiv},
      url={https://arxiv.org/abs/2411.04368},
}

@online{nl2sql-blog,
  author = {Tursio},
  title = {How Far Are NL and SQL in NL2SQL?},
  year = {2025},
  url = {https://www.tursio.ai/blog/how-far-are-nl-and-sql-in-nl2sql},
  lastaccessed = {February 16, 2026},
}

@techreport{woods1972lunar,
  author    = {William A. Woods and Ronald M. Kaplan and Bonnie L. Webber},
  title     = {The Lunar Sciences Natural Language Information System: Final Report},
  institution = {Bolt Beranek and Newman Inc.},
  address   = {Cambridge, Massachusetts},
  number    = {BBN Report 2378},
  year      = {1972}
}

@article{codd1974rendezvous,
  title={RENDEZVOUS Version 1: An Experimental English Language Query Formulation System for Casual Users of Relational Data Bases},
  author={E. F. Codd and
                  Robert S. Arnold and
                  Jean{-}Marc Cadiou and
                  Chin{-}Liang Chang and
                  Nick Roussopoulos},
  journal={Research Report / RJ / IBM / San Jose, California},
  year={1978},
  volume={RJ2144},
}










\end{document}